\documentclass[twocolumn,times]{aastex631} %for arXiv submission
\pdfoutput=1 %for arXiv submission

\usepackage{amsmath}
\usepackage{cases}
\usepackage{soul}

\newcommand{\MHz}{\ensuremath{\, {\rm MHz}}}

\newcommand{\Jy}{\ensuremath{\,{\rm Jy}}}
\newcommand{\K}{\ensuremath{\,{\rm K}}}

\newcommand{\reffg}[1]{Fig.~\ref{#1}}
\newcommand{\reftb}[1]{Table~\ref{#1}}
\newcommand{\refeq}[1]{Eq.~(\ref{#1})}
\newcommand{\refsc}[1]{Sec.~\ref{#1}}

\definecolor{ys}{rgb}{0.0, 0.6, 0.0}

\begin{document}

\title{Searching for axion dark matter gegenschein of the Vela supernova remnant with FAST}

% \author{Wenxiu Yang}\thanks{Contact e-mail: \href{wxyang@nao.cas.cn}{wxyang@nao.cas.cn}}
\author[0009-0006-2521-025X]{Wenxiu Yang}
\affiliation{National Astronomical Observatories, Chinese Academy of Sciences, Beijing 100101, China}
\affiliation{School of Astronomy and Space Science, University of Chinese Academy of Sciences, Beijing 100049, China}

\author{Yitian Sun}
\affiliation{Department of Physics, McGill University, Montr\'{e}al, QC H3A 2T8, Canada}
\affiliation{Trottier Space Institute at McGill, Montr\'{e}al, QC H3A 2T8, Canada}
\affiliation{MIT Center for Theoretical Physics, Massachusetts Institute of Technology, Cambridge, MA 02139, USA}

\correspondingauthor{Yougang Wang}
\author[0000-0003-0631-568X]{Yougang Wang}
\email{wangyg@bao.ac.cn}
\affiliation{National Astronomical Observatories, Chinese Academy of Sciences, Beijing 100101, China}
\affiliation{Key Laboratory of Radio Astronomy and Technology, Chinese Academy of Sciences, A20 Datun Road, Chaoyang District, Beijing 100101, China}
\affiliation{School of Astronomy and Space Science, University of Chinese Academy of Sciences, Beijing 100049, China}
\affiliation{Liaoning Key Laboratory of Cosmology and Astrophysics,  College of Sciences, Northeastern University, Shenyang 110819, China}

\author{Katelin Schutz}
\correspondingauthor{Katelin Schutz}
\email{katelin.schutz@mcgill.ca}
\affiliation{Department of Physics, McGill University, Montr\'{e}al, QC H3A 2T8, Canada}
\affiliation{Trottier Space Institute at McGill, Montr\'{e}al, QC H3A 2T8, Canada}

\author[0000-0003-1962-2013]{Yichao Li}
\affiliation{Liaoning Key Laboratory of Cosmology and Astrophysics,  College of Sciences, Northeastern University, Shenyang 110819, China}

\author{Calvin Leung}
\affiliation{UC Berkeley Department of Astronomy, Berkeley, CA 94720, USA}
\affiliation{MIT Kavli Institute for Astrophysics and Space Research, Massachusetts Institute of Technology, Cambridge, MA 02139, USA}
\affiliation{Department of Physics, Massachusetts Institute of Technology, Cambridge, MA 02139, USA}
\affiliation{NASA Einstein Fellow}

\author[0000-0002-3108-5591]{Wenkai Hu}
\affiliation{Department of Physics and Astronomy, University of the Western Cape, Robert Sobukhwe Road, Bellville, 7535, South Africa}
\affiliation{ARC Centre of Excellence for All Sky Astrophysics in 3 Dimensions (ASTRO 3D), Australia}

\author[0009-0004-8919-7088]{Shuanghao Shu}
\affiliation{National Astronomical Observatories, Chinese Academy of Sciences, Beijing 100101, China}
\affiliation{School of Astronomy and Space Science, University of Chinese Academy of Sciences, Beijing 100049, China}

\author{Kiyoshi Masui}
\affiliation{MIT Kavli Institute for Astrophysics and Space Research, Massachusetts Institute of Technology, Cambridge, MA 02139, USA}
\affiliation{Department of Physics, Massachusetts Institute of Technology, Cambridge, MA 02139, USA}

\correspondingauthor{Xuelei Chen}
\author[0000-0001-6475-8863]{Xuelei Chen}
\email{xuelei@cosmology.bao.ac.cn}
\affiliation{National Astronomical Observatories, Chinese Academy of Sciences, Beijing 100101, China}
\affiliation{Liaoning Key Laboratory of Cosmology and Astrophysics,  College of Sciences, Northeastern University, Shenyang 110819, China}
\affiliation{Key Laboratory of Radio Astronomy and Technology, Chinese Academy of Sciences, A20 Datun Road, Chaoyang District, Beijing 100101, China}
\affiliation{School of Astronomy and Space Science, University of Chinese Academy of Sciences, Beijing 100049, China}

\begin{abstract}
\noindent
Axions are one of the leading dark matter candidates. If we are embedded in a Milky Way dark matter halo comprised of axions, their stimulated decay would enable us to observe a counterimage (``axion gegenschein") with a frequency equal to half the axion mass in the opposite direction of a bright radio source. This spectral line emission will be broadened to {$\Delta \nu/\nu \sim \sigma_d/c \sim 10^{-3}$ 
due to the velocity dispersion of dark matter, $\sigma_d$. In this pilot study, we perform the first search for the expected axion gegenschein image of Vela supernova remnant (SNR) with 26.4 hours of effective ON-OFF data from the Five-hundred-meter Aperture Spherical radio Telescope (FAST) L-band (1.0 - 1.5~GHz) 19-beam receiver. 
Our null detection limits the axion-photon coupling strength to be $g_{a\gamma\gamma} \lesssim 2 \times 10^{-10} \mathrm{GeV}^{-1}$ in the mass ranges of $8.7\,\mu\mathrm{eV} \leq m_a \leq 9.44\,\mu\mathrm{eV}$ and $10.85\,\mu\mathrm{eV} \leq m_a \leq 12.01\,\mu\mathrm{eV} $. 
These results provide a stronger constraint on $g_{a\gamma\gamma}$ in this axion mass range than the current limits obtained by the direct search of axion decay signal from galaxy clusters which uses FAST observations, but is a factor of $\sim 3$ times weaker than the current CAST limit.
Based on our observation strategy, data processing methods, and results, the expected sensitivity will reach $\sim 10^{-11}\mathrm{GeV}^{-1}$ with $\sim 2000$ hours of observation
in the future. 
}
\end{abstract}


\section{Introduction} \label{sec:intro}

Dark matter plays an important role in the composition and evolution of the universe according to the $\Lambda\mathrm{CDM}$ model \citep{2020A&A...641A...6P}. Many theories and experiments have been put forward to reveal the nature of dark matter (see \citealt{2010ARA&A..48..495F} for a review), like for Weakly Interacting Massive Particles~(WIMP) \citep[e.g.][]{1996PhR...267..195J,Steigman_2012,2018RPPh...81f6201R, 2023PhRvD.107j3011G,2023PhRvL.131d1002A,2023PhRvL.131d1003A}, axions \citep[e.g.][]{PhysRevLett.38.1440,2016PhR...643....1M,2021RvMP...93a5004S}, sterile neutrinos \citep[e.g.][]{1994PhRvL..72...17D,2009ARNPS..59..191B,2017PhR...711....1A,2019PhRvD..99h3005N,2021PhRvL.127e1101F}, dark photons\citep[e.g.][]{HOLDOM1986196,2020arXiv200501515F,2021PhRvD.104i5029C,2023PhRvL.130r1001A} and so on. 

In particular, the axion, a pseudo-scalar boson, is one of the most well-studied dark matter candidates arising in Peccei–Quinn theory to solve the strong-CP problem in quantum chromodynamics~(QCD) \citep{PhysRevLett.38.1440}. The interaction of axions and electromagnetic fields enables us to detect them non-gravitationally, with the corresponding Lagrangian term written as $\mathcal{L} = g_{a\gamma\gamma}a\mathbf{E}\cdot\mathbf{B}$, where $g_{a\gamma\gamma}$ is axion-photon coupling strength, $a$, $\mathbf{E}$ and $\mathbf{B}$ are axion, electric and magnetic fields, respectively \citep{PhysRevLett.51.1415}. 

Many existing experiments are designed for laboratory-based detection of axions and axion-like particles \citep{2015ARNPS..65..485G}. For example,  ``haloscopes'' aim to detect the signal of axions in the Milky Way halo converted to photons in a resonant cavity \citep{PhysRevLett.51.1415, PhysRevD.32.2988}. In this way, experiments at Rochester–Brookhaven–Fermilab (RBF; \citealt{PhysRevLett.59.839, PhysRevD.40.3153}) and the University of Florida (UF) %experiments  
\citep{PhysRevD.42.1297}, as well as the Axion Dark Matter eXperiment (ADMX; \citealt{PhysRevD.69.011101, PhysRevLett.121.261302}) have provided strong constraints on axion dark matter. 
On the other hand, axions can be produced in the Sun through the Primakoff conversion of plasma photons in electromagnetic
fields \citep{2011JCAP...06..013I}. 
These solar axions can be converted to X-ray photons in powerful magnetic fields, which is the detection strategy used by ``helioscopes'', including the
CERN Axion Solar Telescope~(CAST; \citealt{2017NatPh..13..584A}) and the International AXion Observatory~(IAXO; \citealt{2014JInst...9.5002A, 2019JCAP...06..047A}). 
Additionally, purely laboratory-based detection strategies, such as light-shining-through-wall experiments  
\citep{2011ConPh..52..211R} can constrain the existence of axions without relying on an astrophysical source of axions. 

Axion-like particles can be detected with astronomical observations via axion conversion and decay to Standard Model particles. 
However, the time scale for spontaneous axion decay into two photons is $\tau_{a\gamma\gamma} = 64\pi\hbar/m_a^3 c^6 g_{a\gamma\gamma}^2 $ \citep{2020arXiv200802729G}, where $m_a$ is the axion mass. An estimation with $m_ac^2 \sim 10\mu \mathrm{eV}$ and $g_{a\gamma\gamma} \sim 10^{-10} \mathrm{GeV^{-1}}$ implies a time scale of $\sim 10^{32}$ years, which is significantly longer than the age of the Universe. 
In the presence of external radiation, axions can undergo stimulated decay at a much faster rate, producing 
two nearly back-to-back photons (one moving in the same direction as the incoming radiation, and the other in the opposite direction). There are many astrophysical environments where one can perform a search for the forward radiation produced by this stimulated decay process, 
including dwarf spheroidal galaxies, the Galactic center and galaxy clusters 
\citep[e.g.][]{2019JCAP...03..027C}. Furthermore, in the opposite direction of a bright radio source, we can expect to observe a counter-image produced by stimulated decay of axions in the Milky Way dark matter halo \citep{2020arXiv200802729G, 2022PhRvD.105b3023A, 2022PhRvD.105f3007S, 2022PhRvD.105g5006B, 2023arXiv230906857A, 2023arXiv231003788S}. 
The counter-image is sometimes called ``axion gegenschein'', which refers to a similar effect where sunlight is scattered by dust in the direction opposite the sun. This novel technique provides a possibility for the indirect detection of %non-QCD 
axion-like particles using existing or future telescopes. 

The Five-hundred-meter Aperture Spherical radio Telescope (FAST; \citealt{2011IJMPD..20..989N}) is the most sensitive single-dish radio telescope in the world.  
Previous forecasts \citep{2022PhRvD.105f3007S} have shown that FAST could constrain the axion-photon coupling in new parameter space through the non-observation of an axion gegenschein signal with hundreds of hours integral time. 
FAST is particularly well-suited for the observation of gegenschein induced by a single source compared with array telescopes such as the Square Kilometre Array (SKA; \citealt{2009IEEEP..97.1482D}) and the Canadian Hydrogen Observatory and Radio-transient Detector (CHORD; \citealt{2019clrp.2020...28V}). Notably, The L-band (1.05 - 1.45~GHz) 19-beam receiver can cover a $\sim$ 20 arcmin $\times$ 20 arcmin field of view %sky area 
\citep{2018IMMag..19..112L,2020RAA....20...64J,2020MNRAS.493.5854H}, comparable to the expected angular extent of a gegenschein image. 

\begin{figure}
    \centering
    \includegraphics[width=0.47\textwidth]{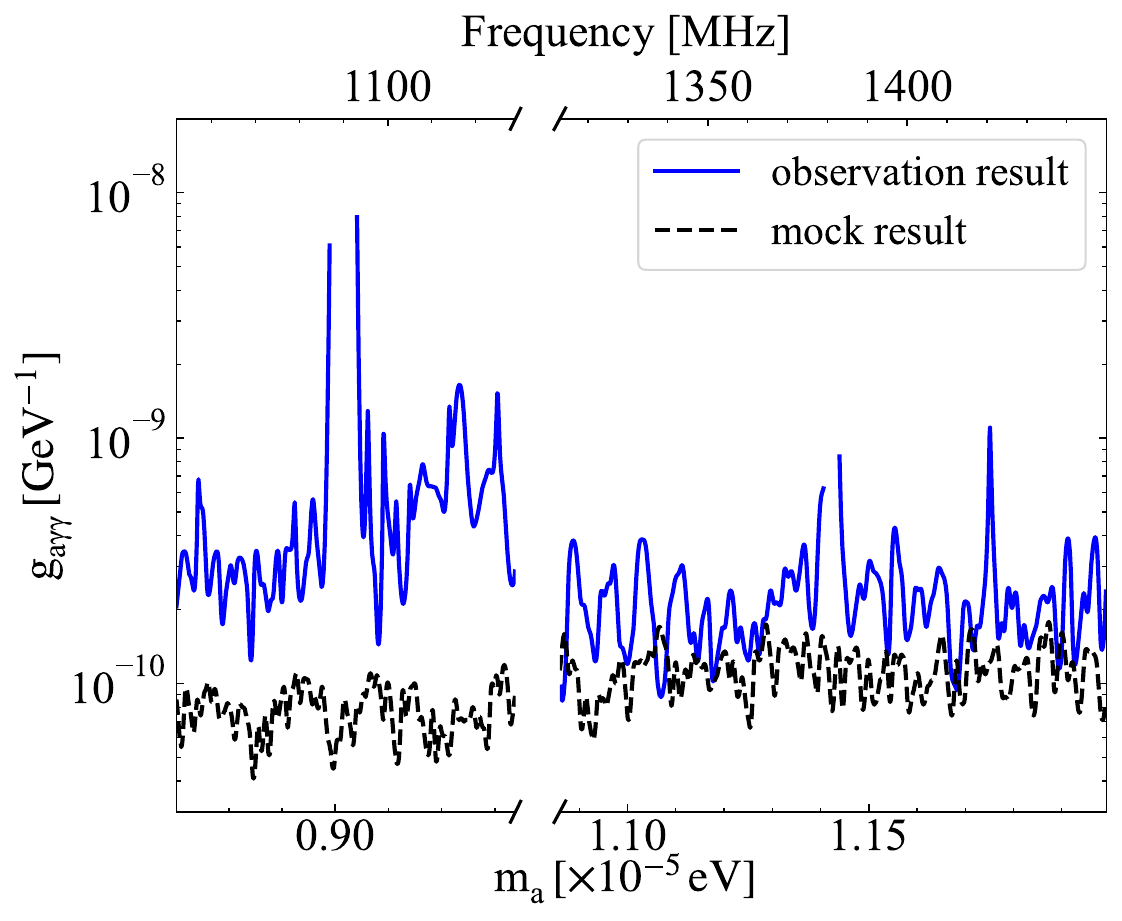}
    \caption{Results of constraint on $g_{a\gamma\gamma}$. The blue solid line displays the upper limits with 95\% C.L., while the black dashed line shows the expected limit for pure thermal noise spectrum with our observation parameters. 
   }
    \label{fig:detect_limit}
\end{figure}

In this work, we perform a pilot study using FAST to search for the a possible gegenschein image sourced by Vela supernova remnant (SNR). 
The non-detection of a convincing axion decay line emission signal allows us to set a strong constraint on the axion-photon coupling strength $g_{a\gamma\gamma}$, as shown in \reffg{fig:detect_limit}. This result is discussed in more details in Sec.\ref{sec:result}.

The rest of this paper is organized as follows. In \refsc{sec:theory}, we introduce the theory of axion gegenschein, as well as the model for a specific source as studied in \citet{2022PhRvD.105f3007S}. Observation data are described in \refsc{sec:obs}, then \refsc{sec:process} is about data processing. We show the results of a search for an axion signal and the resulting parameter constraints in \refsc{sec:result} and provide a detailed discussion in \refsc{sec:discuss}. Finally, our work is summarized in \refsc{sec:summary}.

\section{Theory} \label{sec:theory}

\subsection{Axion gegenschein} \label{subsec:theory_axion}

\begin{figure}
    \centering
    \includegraphics[width=0.47\textwidth]{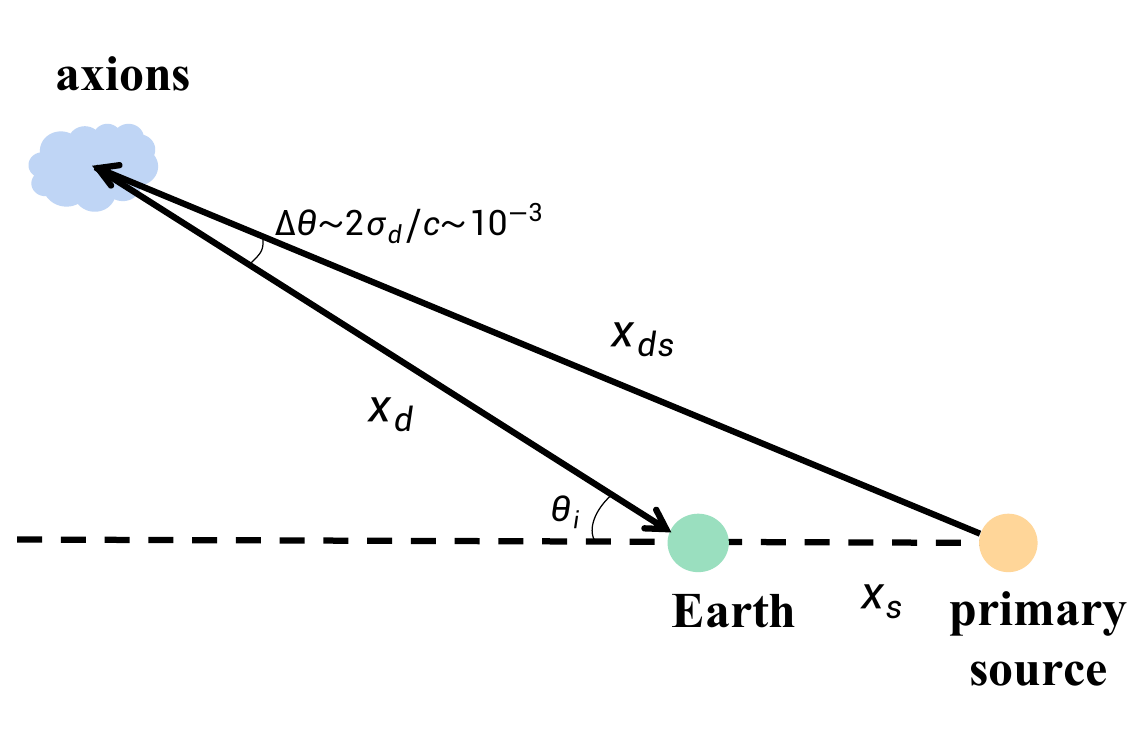}
    \caption{Schematic of axion gegenschein geometry. }
    \label{fig:geometry}
\end{figure}

A schematic diagram of the geometry of the axion gegenschein is shown in \reffg{fig:geometry}. The radiation from a bright primary source passes by and stimulates the decay of axion dark matter particles. The decay photons emitted in this process are nearly back-to-back (moving in the direction of the stimulating radiation and the near-opposite direction) and have a frequency corresponding to half the axion mass. 
Considering the high occupation number of axion dark matter and ignoring the axion back-reaction ({which is very negligible}), the classical field equation describing stimulated axion decay can be expressed as

\begin{equation}\label{eq:classical_field}
    (\partial_t^2 - c^2 \nabla^2) \vec{A_1} = -g_{a\gamma\gamma}(c \vec{\nabla} \times \vec{A_0}) \partial_t a\ ,
\end{equation}

\noindent where $a(t)$ is the background axion field, and $\vec{A_0}$ and $\vec{A_1}$ are the vector potential of the incoming and stimulated electromagnetic radio wave, respectively. As in \citealt{2019PhRvL.123m1804A} and \citealt{2021RvMP...93a5004S}, by Fourier transforming and {treating $\vec{A_1}$ as a perturbation, the flux density of axion gegenschein can be expressed as 

\begin{equation}\label{eq:Sg}
    S_g = \frac{ \hbar c^4 g_{a\gamma\gamma}^2}{16} \int_0^{\frac{t_0c}{2}} S_{\nu}(\nu_a, x_d) \rho(x_d)\mathrm{d}x_d \,,
\end{equation}

\noindent where $S_{\nu}(\nu_a, x_d)$ is the specific flux density at the frequency $ \nu_a = m_ac^2/2h $, $x_d$ is the distance between the observer and the decaying axions, 
and the density $\rho$ of axion dark matter in the Milky Way is assumed to follow the Navarro-Frenk-White profile \citep{1996ApJ...462..563N,1997ApJ...490..493N} 

\begin{equation}\label{eq:rho} 
    \rho(r) = \frac{\rho_0}{\frac{r}{r_s}\cdot\left(1+\frac{r}{r_s}\right)^2}\,,
\end{equation}

\noindent where $r$ is the galactocentric radius and $ r_s = 16~\mathrm{kpc} $ is the scale radius. 
We take the local density as $\rho(r_{\odot}=8.22~\mathrm{kpc}) = 0.46~ \mathrm{GeV/ c^{2}/cm^{3}}$ \citep{2018MNRAS.478.1677S, 2020MNRAS.494.6001N}.  

From \refeq{eq:Sg}, the gegenschein image is a superposition of contributions to the signal coming from different values of $x_d$ along the line of sight. Due to the dark matter velocity dispersion of $ \sigma_d \sim 116\mathrm{km/s} $, the overall image is spatially extended in a way that depends on the finite angular size and distance of the primary source. 
The expected line emission spectrum will also be broadened by the Doppler effect. 

\begin{table} 
    \centering
    \caption{Parameters of Vela SNR.}
    \label{tab:Vela}
    \begin{tabular}{cc}
        \hline
         \quad Parameter \quad & \quad Fiducial Value and {Uncertainty} \\
         \hline
         Position $(l,\, b)$ & ($263.55^{\circ},\,-2.79^{\circ}$) \\
         Distance $x_s$ & $287^{+19}_{-17}$ pc \\
         Age $t_0$ & $1.2\pm 0.3 \times 10^4$ years \\
         MFA time  $t_{\rm MFA}$ & $100^{+200}_{-70}$ years \\
         Spectral index $\alpha$ & $0.74\pm 0.04$ \\
         Electron model $S_{\nu}$ & $S_{\nu} \propto t^{-4p/5}$\\ 
         Alternative electron model & $S_{\nu} \propto t^{-2(p+1)/5} $ \\
         Flux density $S_{\rm 1GHz}$ & 610~Jy \\
         \hline
    \end{tabular}
\end{table}

\subsection{Gegenschein from Vela} \label{subsec:theory_vela}

Based on the analysis in \citet{2022PhRvD.105f3007S}, Vela SNR could be an optimal primary source for stimulating axion decay in the region of the sky that is observable with FAST. In detail, a simplified SNR evolution model indicates that Vela SNR has undergone two main phases of evolution: the free-expansion phase and the Sedov-Taylor phase \citep{2014A&A...561A.139S, 2015A&ARv..23....3D}. The specific luminosity evolution shown in the left plot of \reffg{fig:Vela} is calculated using a model with fiducial values given in Table \ref{tab:Vela} \citep{2003ApJ...596.1137D, 2004A&A...427..525B, 2014A&A...561A.139S}. {Note that the SNR was probably even brighter than indicated in \reffg{fig:Vela} 
during the earliest parts of the free-expansion phase \citep{2021ApJ...908...75B}, and here we conservatively set it to be a constant due to the large modeling uncertainties associated with this epoch.} The luminosity evolution indicates that the SNR was much brighter in the past than it is now, which corresponds to a more luminous gegenschein image at larger values of $x_d$ due to the finite time of flight of the primary radiation and gegenschein signal. 
The flux of the gegenschein signal can be rewritten with the evolution model of Vela SNR analyzed in \citet{2022PhRvD.105f3007S} as 
\begin{align}\label{eq:Sg_vela}
     S_g = & A_g S_{\nu,0}(\nu_a)  \int_{0}^{\frac{(t_0-t_{\rm MFA}) c}{2}} \left(\frac{t_0-\frac{2x_d}{c}}{t_0}\right)^{-\frac{4p}{5}} \rho(x_d)\mathrm{d}x_d \notag\\
     & + A_g S_{\nu,0}(\nu_a) \int_{\frac{(t_0-t_{\rm MFA}) c}{2}}^{t_0c/2} \left(\frac{t_{\rm MFA}}{t_0} \right)^{-\frac{4p}{5}} \rho(x_d)\mathrm{d}x_d 
\end{align}

\noindent where $A_g = \hbar c^4 g_{a\gamma\gamma}^2/16$ is a coefficient related to $g_{a\gamma\gamma}$, $t_0$ is the age of Vela SNR, $t_{\rm MFA}$ is the timescale associated with the onset of magnetic field amplification (MFA), and $p$ is the electron spectral %power law 
index, related to the observed synchrotron spectral index via $\alpha \equiv (p-1)/2$ \citep{2014A&A...561A.139S}. 
The red line in the right figure of \reffg{fig:Vela} shows the flux density at distance $x_d$, while the blue line shows the dark matter density at the corresponding location.

\begin{figure*}
    \centering
    \includegraphics[width=0.49\textwidth]{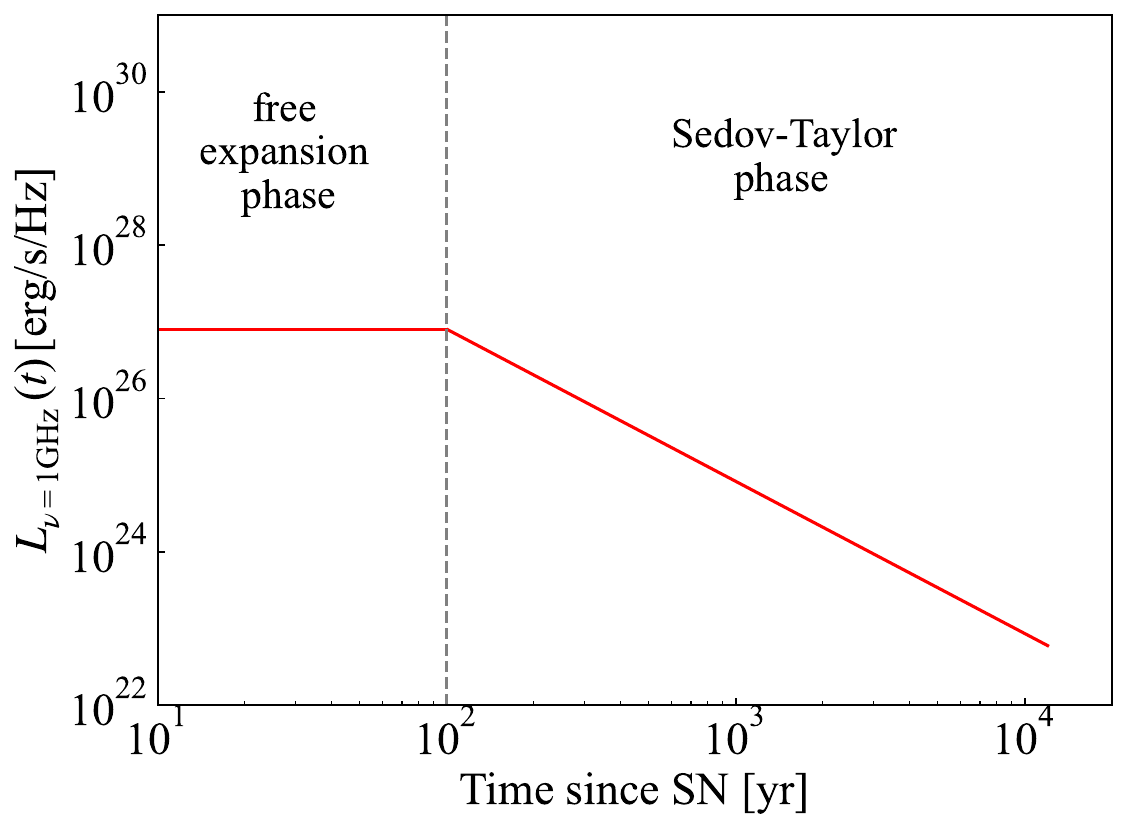}
    \includegraphics[width=0.49\textwidth]{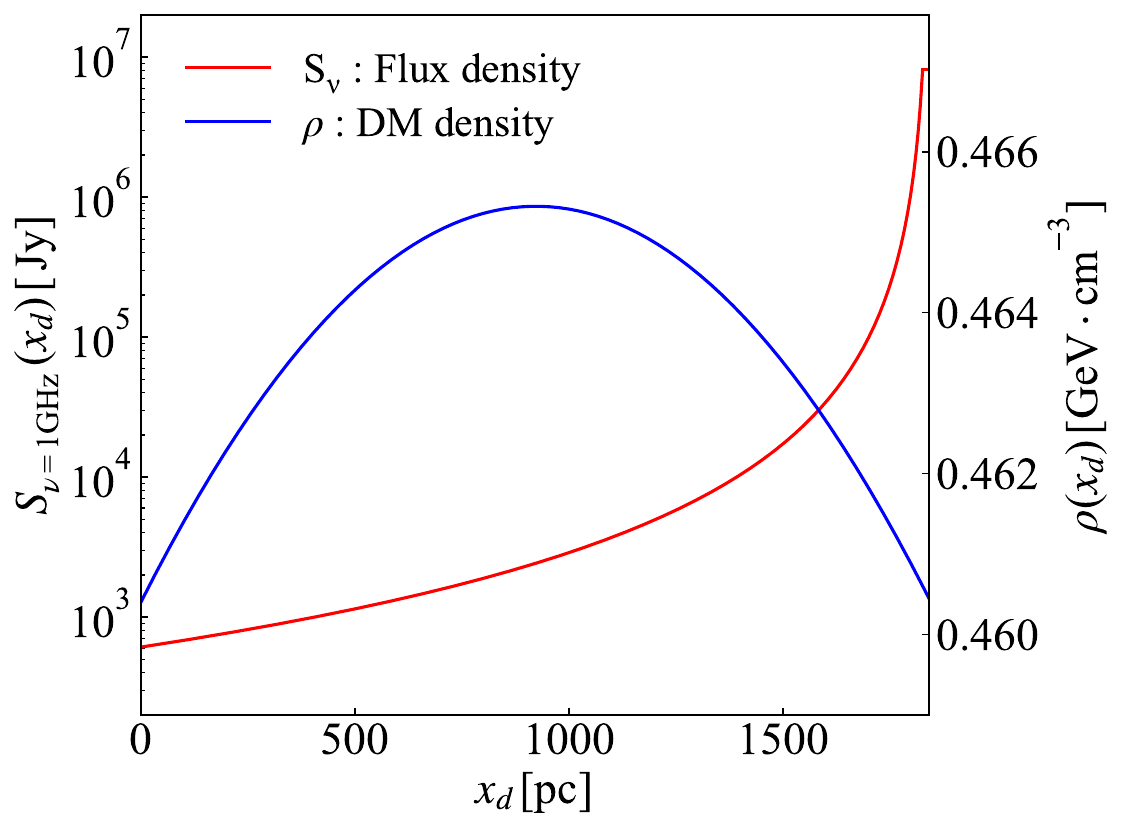}
    \caption{Left: the luminosity evolution model of Vela SNR at 1 GHz. The vertical dashed line marks the transition position of two evolution phases. Right: specific flux density of Vela SNR at 1GHz (red line, left y-axis) and dark matter density (blue line, right y-axis) at different values of $x_d$ along the direction of the gegenschein image. Parameters adopted here are fiducial values in \reftb{tab:Vela}. }
    \label{fig:Vela}
\end{figure*}

The proximity between Vela SNR and Earth leads to a significant angular enlargement of the gegenschein image by a factor of $ x_{ds} / x_s $ as defined in \reffg{fig:geometry} in the small-angle limit. This enlargement enhances the intrinsic spatial smearing 
$ \Delta\theta = 2\sigma_d/c \sim 2.6~ \rm{arcmin}$ caused by dark matter velocity dispersion, such that the total angular extent is $\theta_i = \Delta\theta \cdot x_{ds}/x_s$. 
These geometric considerations are the dominant factor that determines the spatial extent of the axion gegenschein image, rather than the angular distortion sourced by proper motion \citep{2022PhRvD.105f3007S}. The intensity $I_g$ at each distance $x_d$ and different direction $\hat{n}$ can be inferred given $S_g(x_d)$ through

\begin{equation}\label{Sv_xd} 
    S_g(x_d) = \int I_g(x_d,\hat{n}) d\Omega \,,
\end{equation}

\noindent assuming the axion gegenschein image at $x_d$ follows a 2D Gaussian profile. For simplicity, we also use the flat sky approximation for the small sky patch of the gegenschein image, 
The observed signal flux can be written as 
\begin{equation}\label{eq:S_obs}
    S_{\rm obs} = \int I_g(\hat{n})b(\hat{n})\mathrm{d}\Omega \,. 
\end{equation}
The beam pattern $b(\hat{n})$ adopted here is simplified as a 2D Gaussian profile with full-width half-maximum (FWHM) given in \citet{2020RAA....20...64J} that varies with frequency. 

The normalized intensity of the axion gegenschein image induced by Vela SNR is shown in \reffg{fig:src_beam}. 
For comparison, the 19 L-band beams of FAST are indicated by the black circles, which correspond to the FWHM at 1.0~GHz.  
Notably, the bright central part of the predicted gegenschein image is well distributed over the 19 beams.

\begin{table} 
    \centering
 \caption{Observation parameters.}
     \label{tb:obs}
    \begin{tabular}{cccc}
        \\
        \hline
         \quad Dataset \quad & \quad Year \quad & \quad OFF-source \quad & \quad Observation \quad \\ 
          &  \quad & \quad Position \quad & \quad Time \\
         \hline
         
         \quad A \quad & \quad 2021 \quad & \quad 1 \quad & \quad 20h ON \\
         \quad   \quad & \quad   \quad & \quad   \quad & \quad + 100min OFF \\ \hline
         
        \quad B \quad & \quad 2022 \quad & \quad 1 \quad & \quad 190min ON \\ 
        \quad   \quad & \quad   \quad & \quad   \quad & \quad + 190min OFF \\ 
        \quad & \quad      \quad & \quad 2 \quad & \quad 130min ON \\ 
        \quad & \quad      \quad & \quad   \quad & \quad + 130min OFF \\ \hline
           
         \quad C \quad & \quad 2023 \quad & \quad 3 \quad & \quad 144min ON \\
         \quad   \quad & \quad   \quad & \quad   \quad & \quad + 144min OFF \\
        \quad & \quad      \quad & \quad 4 \quad & \quad 144min ON \\
        \quad & \quad      \quad & \quad  \quad & \quad + 144min OFF \\
         \hline
    \end{tabular}
\end{table}

\section{Observations} \label{sec:obs}

We obtained a total of 59 hours of observation time 
through our proposal in 2021 (PT2021\_0020) and 2022 (PT2022\_0185), including 30 hours of Grade A (i.e. high priority) time and 15 hours of Grade B time from proposal 2021, and 14 hours of Grade B time from proposal 2022. This includes time for target tracking, OFF-source point tracking, calibration, and telescope slewing. We carried out three observation sessions in the summer of 2021, June 2022, and June 2023, which we will refer to as data A, data B, and data C respectively. For simplicity, hereafter we will use the uppercase ``ON" and ``OFF" to denote the data at ON-source and OFF-source points, whereas we will use the lowercase ``on" and ``off" to distinguish data with and without noise injection. The observation is carried out with the 19-beam L-band receiver, with the frequency range of 1050-1450 MHz. We tested several different observation modes. The time-frequency (waterfall) plot of a sample of raw data is shown in Figure \ref{fig:obs_wfp}, with the left panel for the data A, middle panel for the data B, and right panel for the data C. 
The time-averaged spectra of the sample data for the 19 beams are shown in Figure \ref{fig:obs_spec}. Parameters of these observation are listed in \reftb{tb:obs}.

\begin{figure}
    \centering
    \includegraphics[width=0.5\textwidth]{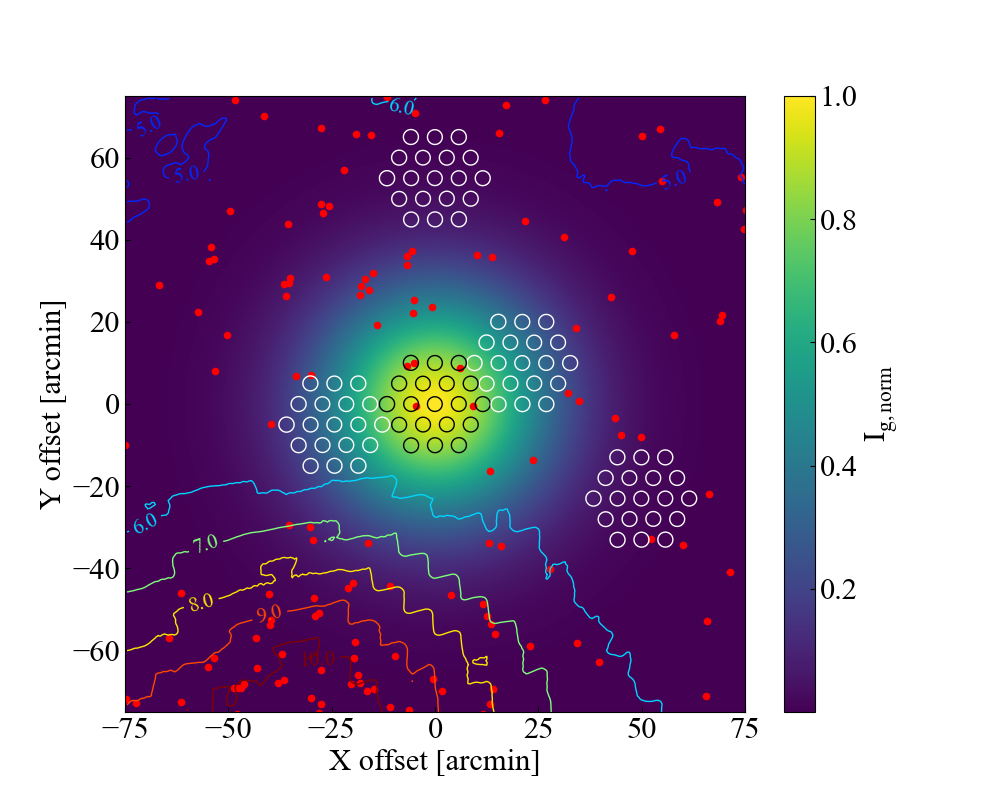}
    \caption{Surroundings of the Vela SNR gegenschein source. The normalized intensity of the gegenschein is shown in color map, the FAST beams are shown as circles at the ON-source (black) and four OFF-source (white) positions. Also shown are galactic radio emission estimated by extrapolating from the Haslam sky map at 408 MHz (counter image, \citealt{1982A&AS...47....1H,2015MNRAS.451.4311R}) and NVSS sources with flux $>10$mJy (red points) at 1GHz.} 
    \label{fig:src_beam}
\end{figure}

\begin{figure*}[ht!]
    \centering
    \includegraphics[width=0.32\textwidth]{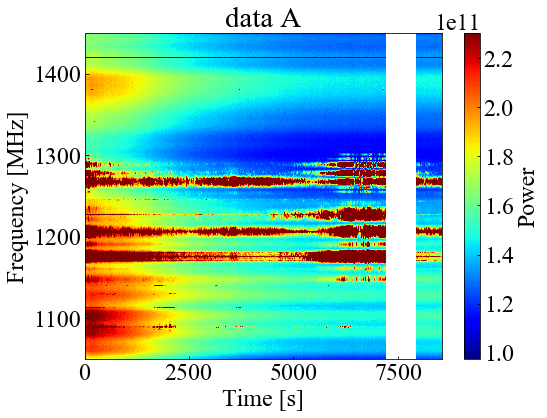}
    \includegraphics[width=0.32\textwidth]{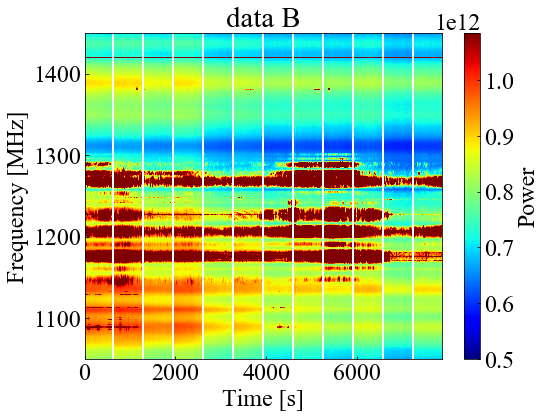}
    \includegraphics[width=0.32\textwidth]{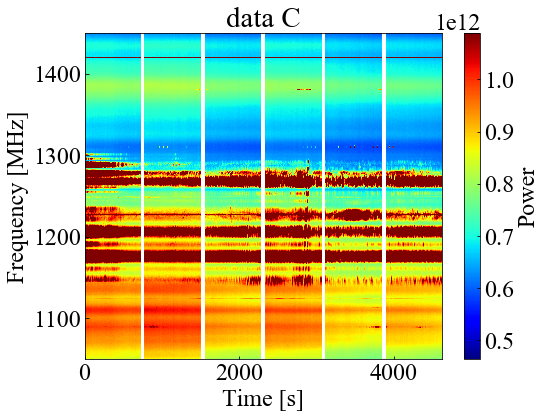}
    \caption{Waterfall plots of raw data of M01, XX polarization. The blank regions correspond to the switching  between ON-source and OFF-source positions. Left: data A, Day 1. Middle: data B, Day 3. Right: data C, Day 1.}
    \label{fig:obs_wfp}
\end{figure*}
\begin{figure*}
    \centering
    \small
    \includegraphics[width=0.32\textwidth]{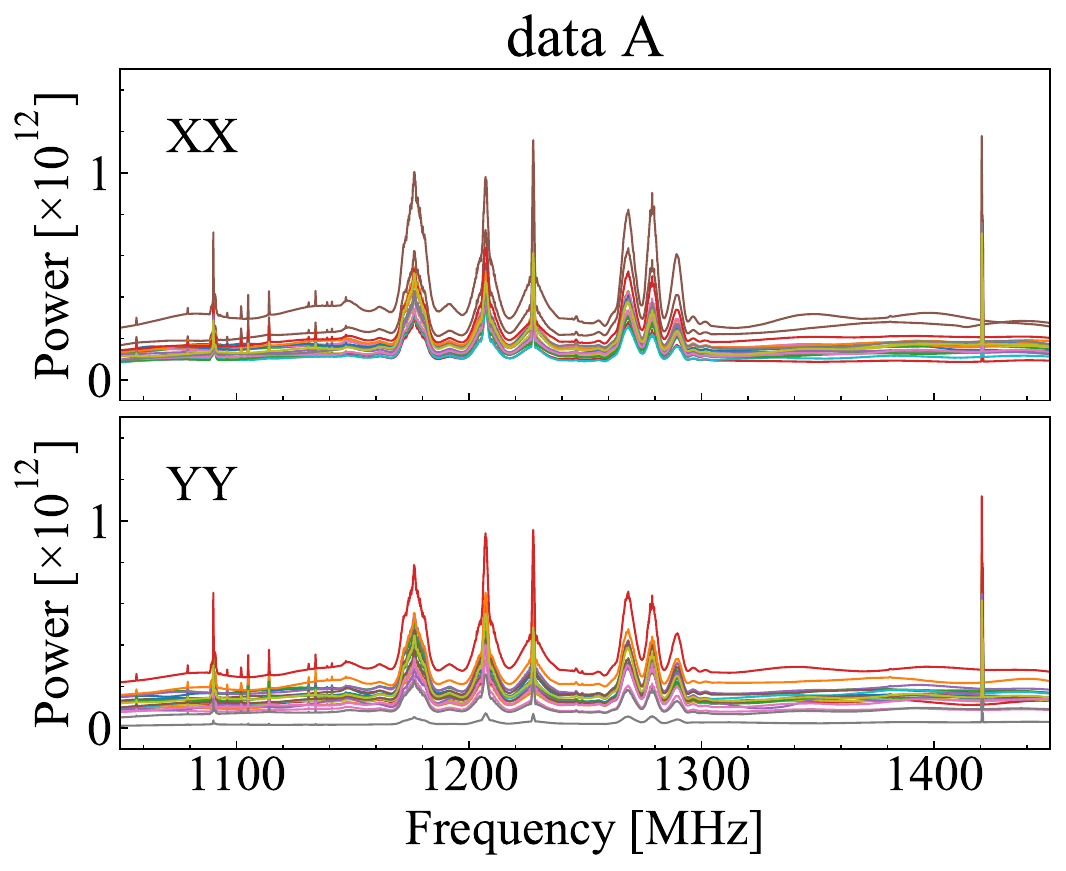}
    \includegraphics[width=0.32\textwidth]{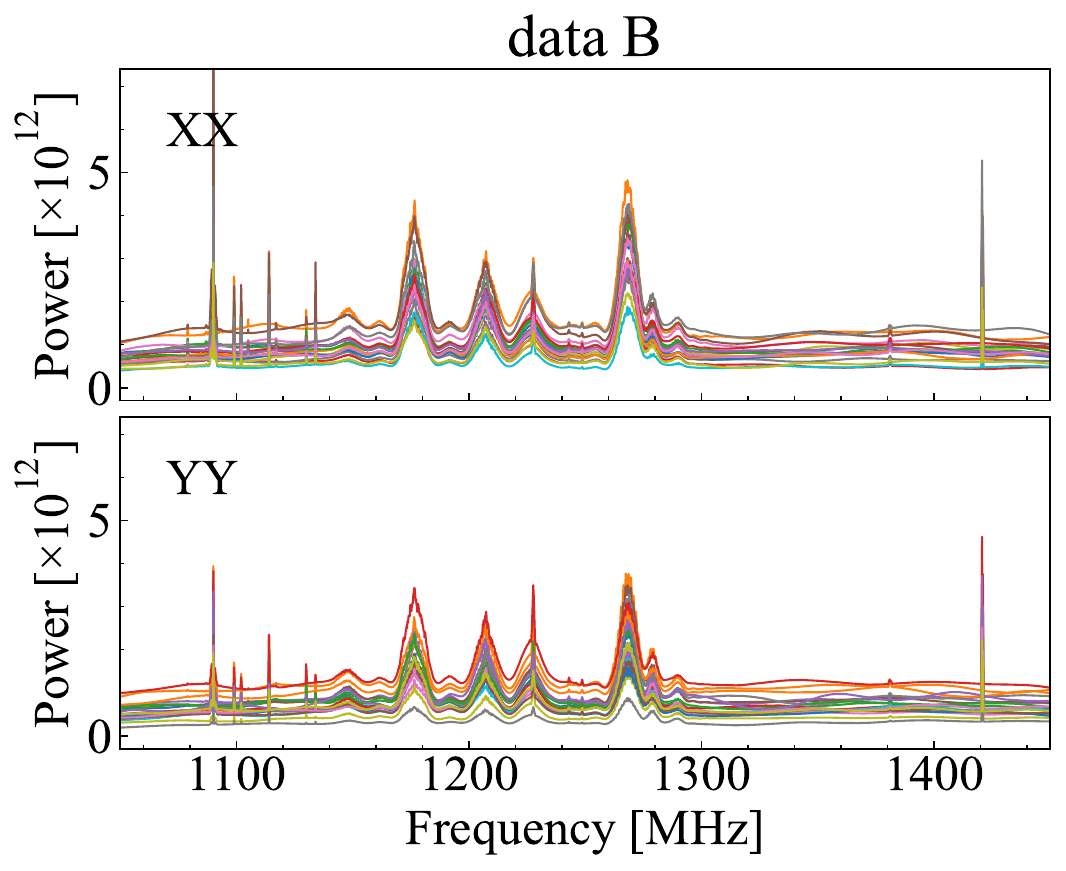}
    \includegraphics[width=0.32\textwidth]{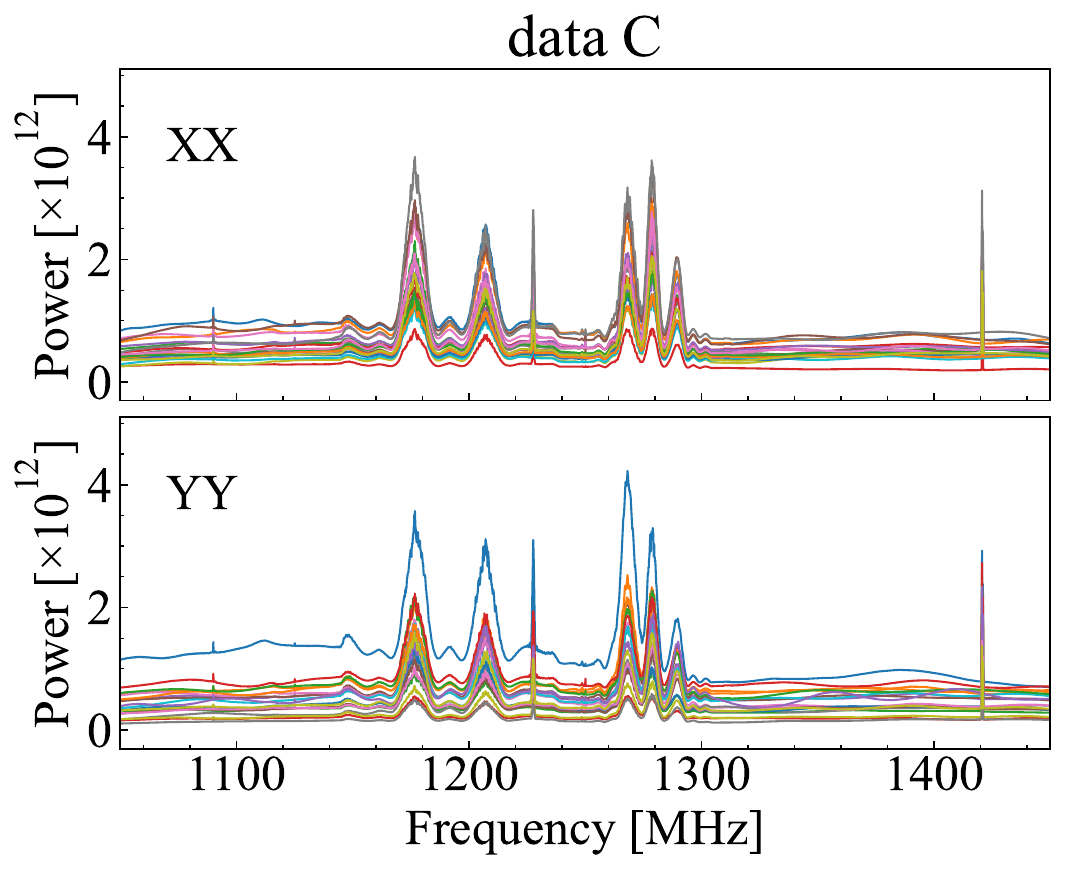}
    \caption{Raw spectra at 1050-1450~MHz in Day 1 of data A (left), Day 3 of data B (middle) and Day 1 of data C (right)} of XX polarization (top) and YY polarization (bottom). Each line represents data from a single beam.
    \label{fig:obs_spec}
\end{figure*}

\subsection{Observation in 2021} \label{subsec:obs_2021}

During the observation in 2021, the 30 hours of time (A) are equally split into 10 nights, each containing 2 hours of tracking observation of the antipodal position for the Vela SNR for its gegenschein, 10 minutes for tracking a nearby OFF-source point, and 20 minutes for a calibrator (3C409) observation with the MultiBeamCalibration mode. The left panel of \reffg{fig:obs_wfp} shows the raw ON- and OFF-source data of a single beam and polarization during one day, while the left panel of \reffg{fig:obs_spec} shows the time-averaged spectra of all 19 beams and 2 polarization channels of the ON-source observation. 

The internal noise diode was used in its low temperature ($\sim$1 K) mode for gain calibration. We used a high-cadence noise injection mode, which had been used in the the Commensal Radio Astronomy FAST Survey (CRAFTS, \citealt{2018IMMag..19..112L}). The noise was injected for 81.92 $\mu$s every 196.304 $\mu$s, which is similar to the pulsar sampling time scale. This mode ensures the temporal continuity of data, which enabled us to perform high time-resolution calibration with the help of the pulsar backend, while simultaneously searching for pulsars and Fast Radio Bursts (FRBs) as byproducts.

\subsection{Observation in 2022} \label{subsec:obs_2022}

Based on a preliminary analysis of the data obtained in 2021, we revised our observation strategy when carrying out the further observation of 15 hours of time (B) in the summer of 2022. In this run, equal time is assigned to the ON- and OFF-source, and we made a switch between them every 10 minutes to reduce the error induced by  bandpass instability. We also used an additional OFF-source point, to better assess the sky background. In the observations of 5 days, we obtained a total of 320 minutes each of ON-source and OFF-source data. The sky calibrator was observed in the same manner as it was for data A.  
Moreover, a more traditional noise injection mode was used for data B,
in which the noise diode was turned on for $\sim$ 1s every $\sim$ 8s to avoid noise overflow in the high-cadence mode. Examples of waterfall plots and spectra from data B are shown in the middle panel of \reffg{fig:obs_wfp} and \reffg{fig:obs_spec}.

\subsection{Observation in 2023} \label{subsec:obs_2023}

In the 2023 observations, we continued to use the same noise injection mode and sky calibrator as those used in 2022. In data C, we selected 2 new OFF-source positions which are further from the ON-source position than before, to ensure that the gegenschein signal would not be subtracted during background subtraction. The 14 hour observation was distributed over 4 days. On each day, we used the ON-OFF mode with both OFF-source positions for better comparison of different OFF-sources, and to minimize the influence of the choice of OFF-source on our result. The switching period for ON- and OFF-source position is 24 minutes as shown in the right panel of \reffg{fig:obs_wfp}. In total, we obtained 9.6 hours of pure ON-OFF mode observation in data C.

\section{Data processing} \label{sec:process}

We divide the raw data into three frequency bands (1050-1129 MHz, 1129-1313 MHz, and 1313-1450 MHz) according to the level of RFI (c.f. \reffg{fig:obs_spec}). Data in the 1129-1313 MHz band are discarded as they are too contaminated by RFI. We also re-bin the raw data from a frequency resolution of $\Delta\nu\sim$7.6kHz to $\Delta\nu\sim$122 kHz to reduce the amount of computation. Tests of sample data show that the result is nearly identical to the one obtained by first flagging RFI then rebinning. We smooth the OFF-source data obtained in 2021 along the frequency axis at the beginning (before rebinning) to lower the level of white noise.

The data processing follows the standard pipeline for a single dish radio telescope (\citealt{2002ASPC..278..293O, 2023ApJ...954..139L}), the main steps are:\\ 
\begin{enumerate}
    \item temporal RFI flagging;
    \item receiver gain and bandpass calibration using the noise diode;
    \item absolute flux calibration using the sky calibrator;
    \item baseline and standing wave removal;
    \item determining the detection limit and performing a signal search.
\end{enumerate}

\subsection{Temporal RFI flagging} \label{subsec:process_RFI}

During the observation, very strong  interference \citep{2020RAA....20...75Z} occasionally occurs at some frequencies (e.g. RFI near 1380~MHz), which contaminates the spectrum. Another source of temporal interference is a jump in the signal level across all frequency bands. We flag these two kinds of temporal interference, while keeping other sources of RFI and strong spectral lines (e.g. galactic HI emission at 1420.4 MHz) because our target signal has the same characteristics. To this end, the temporal flagging process is run on each frequency channel. We flag the outliers deviating over 4.5 standard deviations from the smoothed-band baseline, and iterate the process until no new point is flagged. To reduce residual  interference, adjacent time points are also masked. The percentage of masked time points at each frequency channel ($f_{\rm mask}(\nu)$) is shown in the right figure of \reffg{fig:process_RFI1d}. We can see the higher frequency band (1313-1450\MHz) is clean except for a few contaminated channels near 1380\MHz. For the lower frequency band, the data is more influenced by RFIs. 

\begin{figure*}
    \centering
    \small
    \includegraphics[width=0.39\textwidth]{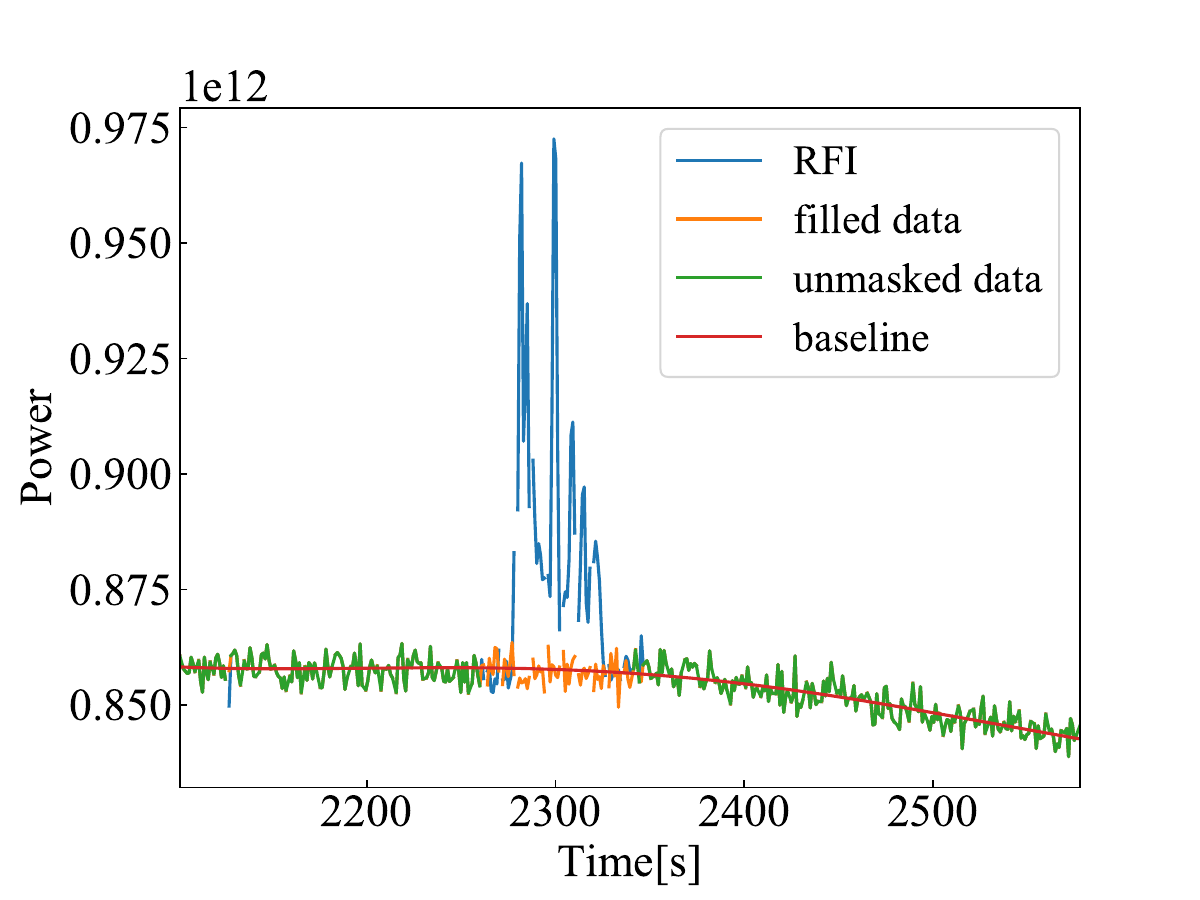}
    \includegraphics[width=0.6\textwidth]{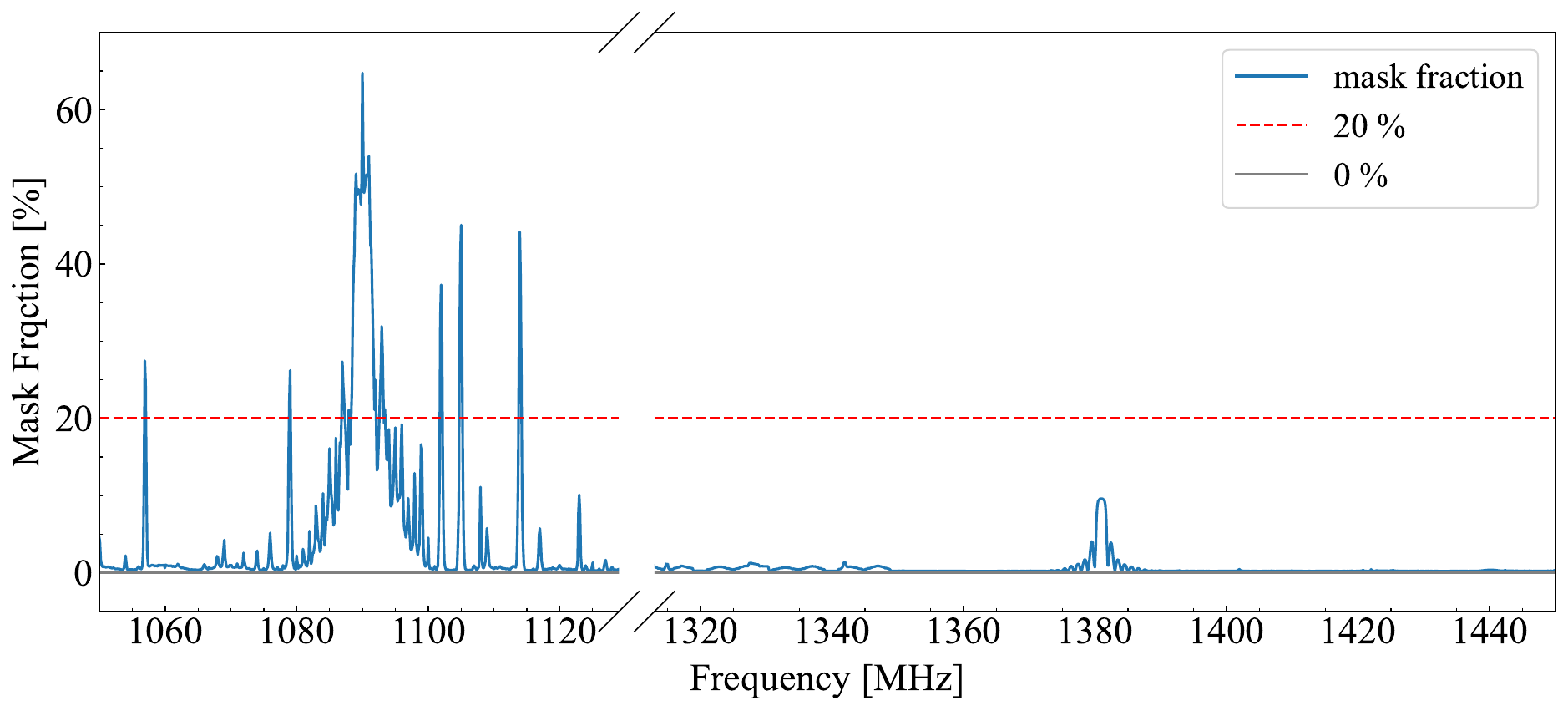}
    \caption{Left: A demonstration of flagging and refilling on a single frequency point (1381.35 MHz). Lines with different colors show data with RFI (blue), data without interference (green), smoothed baseline (red), and refilled data at the contaminated time range (orange), respectively. 
    Right: The percentage of flagged time points at each frequency channel. The red dashed line and gray solid line mark the value of 20\% and 0\% respectively.}
    \label{fig:process_RFI1d}
\end{figure*}

To maintain continuity in the frequency domain, the masked regions are refilled with interpolated data with white noise added as shown in the left plot of \reffg{fig:process_RFI1d}. 
The target signal does not vary rapidly over time, so this interpolating procedure will not introduce false signal. Even so, to be conservative in our search, we still mask frequencies 
whose time variation is mostly contaminated before searching for a signal in the final frequency spectrum. 
Furthermore, to avoid artificially underestimating the noise level of the final spectrum, we introduce a correction factor $C_{\rm mask}(\nu) = \sqrt{1-f_{\rm mask}(\nu)}$ with the mask fraction $f_{\rm mask}(\nu)$ when derive the result at \refsc{sec:result} to recover the noise level without interpolated data.

Certain beam feeds show significant temporal fluctuations which seem to be periodic, as shown in 
\reffg{fig:badfeed} for example. The origin of these fluctuations is unknown. In the subsequent analysis, we do not use such bad data.  

\begin{figure}
    \centering
    \includegraphics[width=0.5\textwidth]{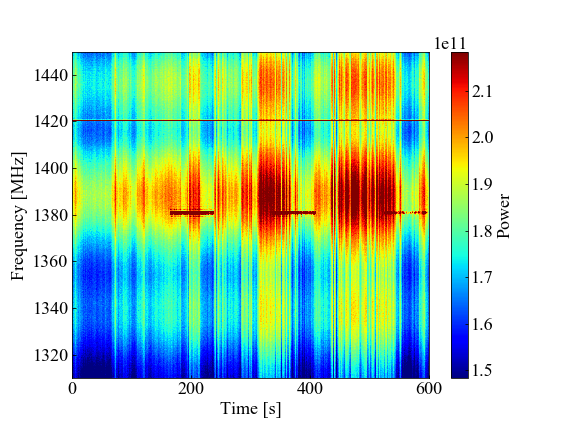}
    \caption{An example of bad data from data A, Day 8, M10, yy polarization}
    \label{fig:badfeed}
\end{figure}

\subsection{Bandpass and gain calibration} \label{subsec:process_bpgt}

As discussed in \refsc{sec:obs}, noise with temperature $\sim 1\K$ was injected periodically for calibration. We calculate the instrument frequency response (bandpass) $\mathcal{B}(\nu)$ for each time block of $\sim$ 10 minutes in data B and C as
\begin{equation}\label{eq:bp}
    \mathcal{B}(\nu) = \frac{\langle V_{\rm on}(\nu, t) - V_{\rm off}(\nu, t) \rangle_t}{T_{\rm ND}(\nu) \cdot t_{\rm on}/t_{\rm samp}}\,,
\end{equation}
where $V_{\rm on}$ and $V_{\rm off}$ are values measured by the FAST spectrum backend when the noise diode is turned on and off respectively, with $t_{\rm on} = 0.9\ \mathrm{s}$, $t_{\rm samp} = 1.00663296\ \mathrm{s}$ during observation, and the noise diode equivalent temperature $T_{\rm ND}$ has been measured using the hot-load method.

For data A, the calculation is a little more complicated. Since the injected noise signal in the high-cadence mode can only be identified by the FAST pulsar backend, but not in the spectral backend, we first extract the data from the pulsar backend, then re-bin them according to the sampling time of the spectrum backend ($\sim 0.2\ \mathrm{s}$). Following \cite{2024arXiv241208173Y}, the bandpass of the pulsar backend $\mathcal{B}_{\rm psr}(\nu)$ can be calculated using \refeq{eq:bp} with the noise-on and noise-off data ($V_{\rm on, psr}$ and $V_{\rm off, psr}$), as well as the noise injection parameters ($t_{\rm on} = 81.92\ \mu\mathrm{s}$ and $t_{\rm samp} = 98.304\ \mu\mathrm{s}$ for this observation). Then we use a coefficient $C(\nu)$ to convert $\mathcal{B}_{\rm psr}$ to $\mathcal{B}_{\rm spec}$, assuming they have similar bandpass shape but differ in their amplitude, 
\begin{align}
    \mathcal{B}_{\rm spec}(\nu) &= \mathcal{B}_{\rm psr}(\nu) \cdot C_{\rm bp}(\nu) \notag \\
    &= \mathcal{B}_{\rm psr}(\nu) \cdot \left\langle \frac{V_{\rm spec}}{(V_{\rm psr, on} + V_{\rm psr, off})/2} \right\rangle_t \,.
\end{align}
Furthermore, assuming a bandpass shape stable in time, the calculation of temporal fluctuation of gain (bandpass amplitude) $g(t)$ for data B and C is similar, but averaged along frequency as
\begin{equation}
    g(t) = \left\langle \frac{V_{\rm spec, on}(\nu, t) - V_{\rm spec, off}(\nu, t)}{\mathcal{B}_{\rm spec}(\nu)} \right\rangle_{\nu}\,,
\end{equation}
while for the data A of high-cadence noise injection mode, the expression becomes
\begin{equation}
    g(t) = \left\langle \frac{V_{\rm psr, on}(\nu, t) - V_{\rm psr, off}(\nu, t)}{\mathcal{B}_{\rm psr}(\nu)} \cdot \frac{C(\nu, t)}{C_{\rm bp}(\nu)}\right\rangle_{\nu}\,.
\end{equation}

After dividing the raw data by the bandpass and $g(t)$, we obtain the antenna temperature $T_{\rm cal}$ as 
\begin{equation}\label{eq:Tcal}
    T_{\rm cal}(\nu, t) = \frac{V_{\rm spec}(\nu, t)}{\mathcal{B}_{\rm spec}(\nu) \cdot g(t)}\,.
\end{equation}
Note that when calculating the bandpass of data A, the ON-source data are divided to $\sim$ 10 minutes blocks to be consistent with the OFF-source data. We also convolve a Hanning window function or use a median filter in several steps to suppress the white noise and smooth the bandpass and gain.

\subsection{Absolute flux calibration} \label{subsec:process_fluxcal}

In each day's observation, we observe 3C409 as a sky calibrator using the MultiBeamCalibration mode for absolute flux calibration. The spectrum of 3C409 (denoted as $S_{\rm 3C409}(\nu)$) is assumed to follow a power law, and it is related to the observed temperature $T_{\rm 3C409}(\nu)$ by 
\begin{equation}
T_{\rm 3C409}(\nu)= \eta(\nu,\theta_{\rm ZA,0})\cdot G_0
 B(r) S_{\rm 3C409}(\nu),
\end{equation}
where $G_0 = \frac{\pi R^2}{2k_{\rm B}} = 25.6\,\mathrm{K/Jy}$ is the theoretical antenna flux gain, $R=150\ {\rm m}$ is the effective illuminated half size of the FAST,  $\eta(\nu, \theta_{\rm ZA,0})$ is the aperture efficiency at the zenith angle $\theta_{\rm ZA,0}$, $B(r)$ is the beam pattern, which is assumed to be a Gaussian profile with the FWHM given in \citet{2020RAA....20...64J}, while $r$ is the great-circle distance between the pointed direction and the 3C409 position. We use the flux value at 750 MHz and 1400 MHz \footnote{\url{http://ned.ipac.caltech.edu/}} to fit the spectrum parameters, then use this equation to determine the aperture efficiency. To reduce the influence of the different zenith angles of the target source and the calibrator,  we first calculate a correction factor $C_{\eta}(\nu, \theta_{\rm ZA,0})$ at the zenith angles when tracking 3C409, to quantify the difference between our results and those in \citet{2020RAA....20...64J}. Then we calculate $\eta$ for any other zenith angle $\theta_{\rm ZA}$, by multiplying the correction factor on earlier observed results of FAST at the corresponding zenith angle (denoted as $\eta_{\mathrm{Jiang}}(\nu, \theta_{\rm ZA})$),
\begin{align}
    \eta(\nu,t) &= \eta(\nu,\theta_{\rm ZA}(t)) \notag \\
    &= C_{\eta}(\nu, \theta_{\rm ZA,0}) \cdot \eta_{\mathrm{Jiang}}(\nu, \theta_{\rm ZA}(t)).
\end{align}
The antenna temperature is then converted to flux units by dividing $\eta$ and the theoretical flux gain $G_0$, 
\begin{equation}
    S(\nu, t) =  \frac{T_{\rm cal}(\nu, t)}{\eta(\nu, t) \cdot G_0}\,.
\end{equation}

\subsection{Baseline and standing wave removal} \label{subsec:process_rmbslsw}

\begin{figure*}
    \centering
    \includegraphics[width=0.325\textwidth]{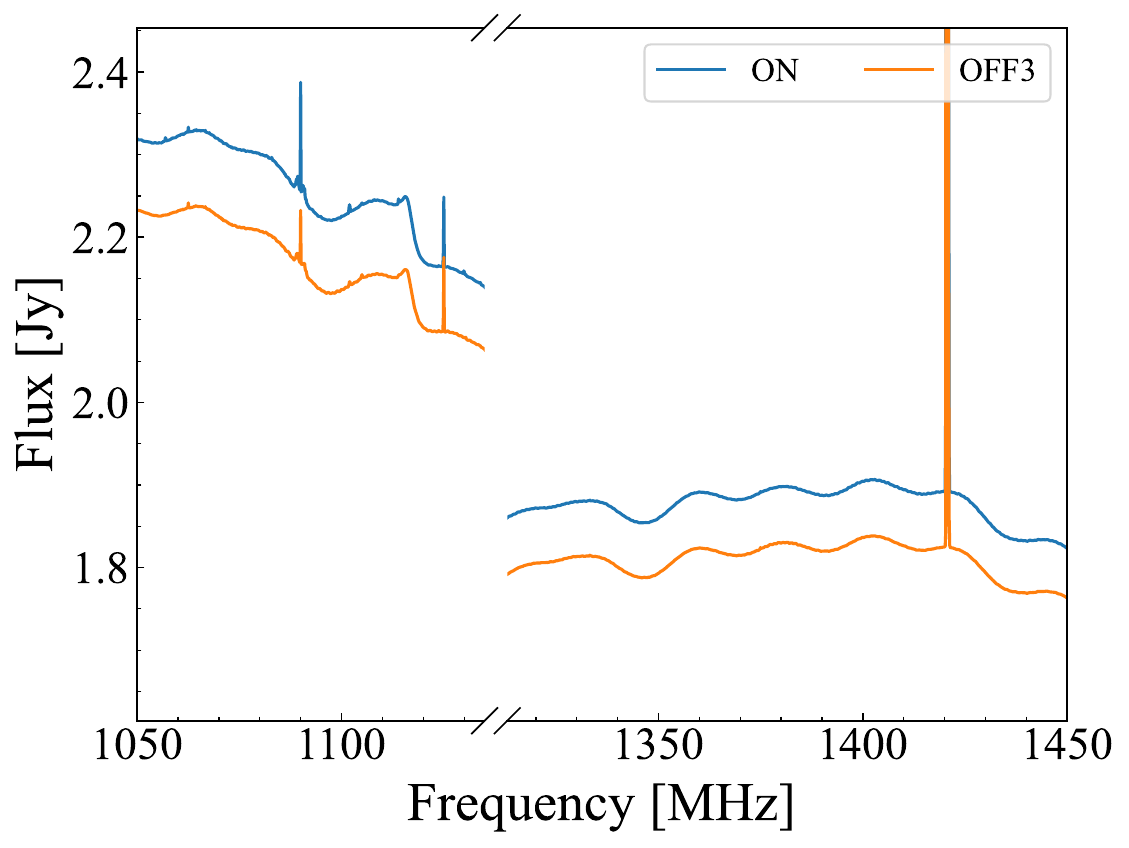}
    \includegraphics[width=0.325\textwidth]{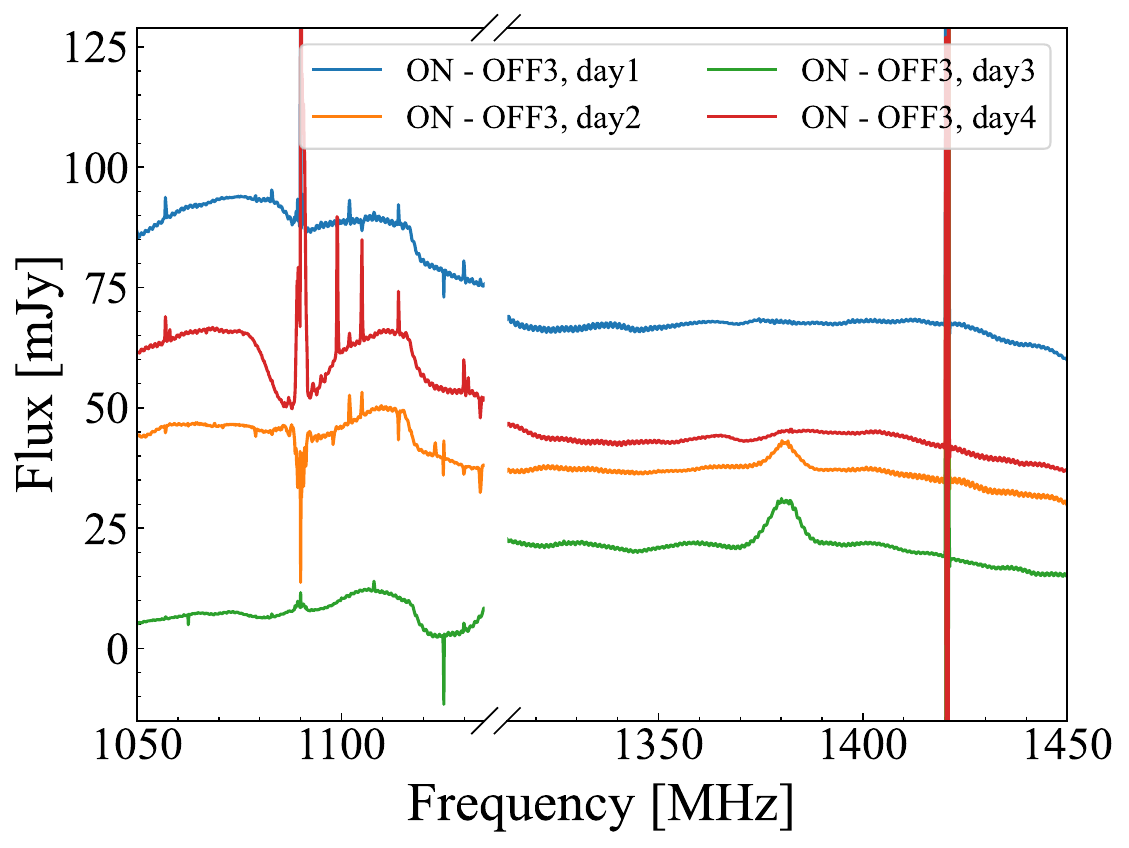}
    \includegraphics[width=0.325\textwidth]{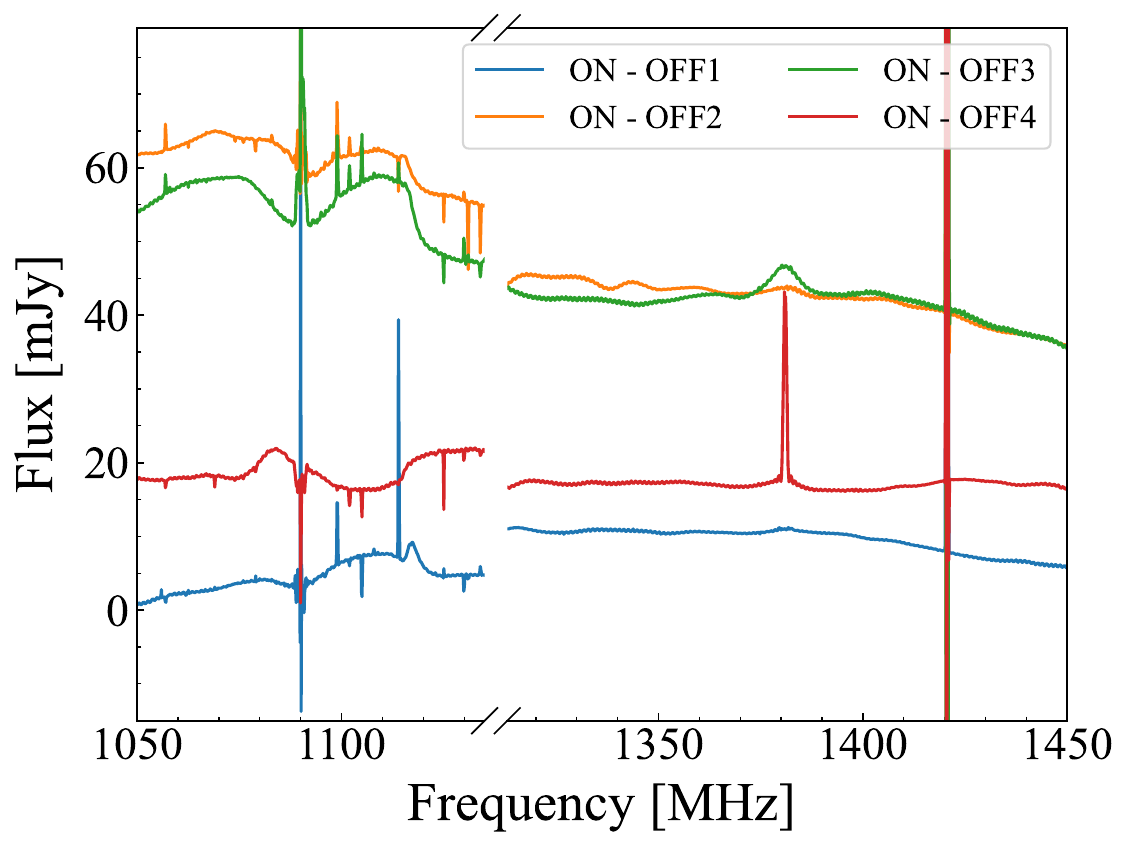}
    \caption{Left: comparison between spectra of $\rm ON$ and $\rm OFF$ from 1 day's observation. Middle: comparison between spectra of $\rm ON-OFF_3$ from 4 different days. Right: comparison between spectra of $\rm ON-OFF_i, i=1,2,3,4$ for 4 days of data.}
    \label{fig:compare_ONOFF}
\end{figure*}

After the calibration, we make an $\rm{ON-OFF}$ subtraction of the time-averaged data in each day, to remove the sky background and the system baseline. An example spectrum of ON and OFF from 1 day of observation is shown in the left panel of \reffg{fig:compare_ONOFF}, and the $\rm ON-OFF_3$ spectra derived for a few days are shown in the middle panel of \reffg{fig:compare_ONOFF}. The amplitude of difference spectra is significantly reduced compared with the raw spectra, because the ON and OFF spectra are very similar. Nonetheless, the flux is not zero, and in fact varies significantly on different days. For the higher frequency (1300--1450 MHz) band, the residue is relatively flat (though there is a bump near 1390 MHz for day 2 and day 3, which could be RFI). Such a spectrum may arise due to different values of the background intensity at the ON position and $\rm OFF_3$ position, and the variation in overall amplitude may reflect the variation in the gain and baseline (details are discussed in \refsc{subsec:discuss_cal}). 
For the lower frequency band (1050--1150 MHz), aside from differences in the overall amplitude, there are also variations in the shape of the band, which may be due to RFI contamination.

We also show the $\rm ON-OFF_i(i=1,2,3,4) $ spectra for the different OFF regions in the right panel of \reffg{fig:compare_ONOFF}, with all days of data combined by averaging. 
Here again we see there are significant differences, especially in the lower frequency band. Also, the spectra for the pair $\rm ON-OFF_1$ and $\rm ON-OFF_2$ are more similar to each other, while the $\rm ON-OFF_3$ and $\rm ON-OFF_4$ spectra are more similar to each other.
These spectral variations limit our ability to detect spectral features in the observation.

\begin{figure*}
    \centering
    \includegraphics[width=0.9\textwidth]{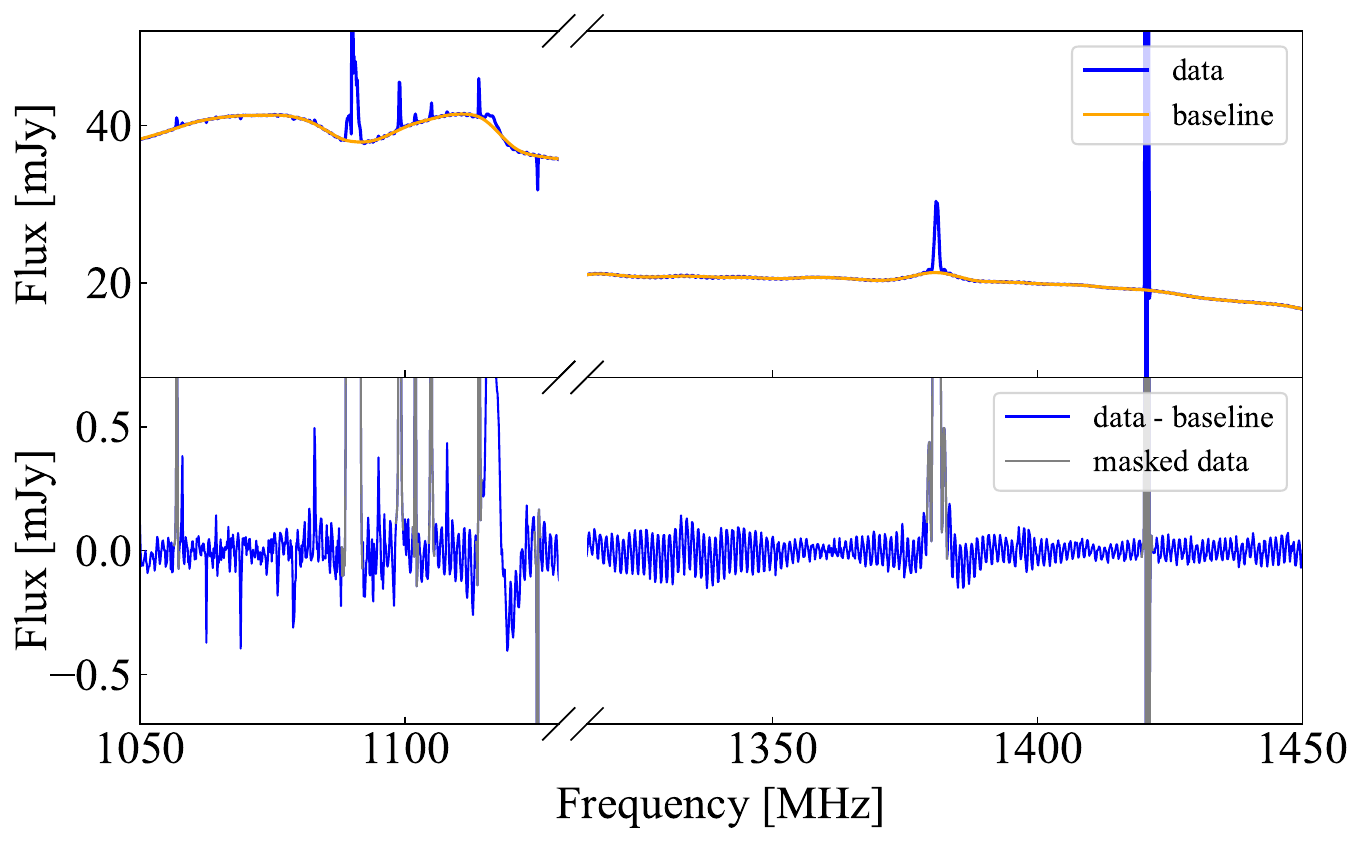}
    \caption{Time-averaged calibrated data at 1050-1135 MHz and 1313-1450 MHz. Top panel: spectrum after flux calibration (blue) and baseline fitting (orange); Bottom panel: baseline subtraction result.  }
    \label{fig:process_rmbslsw}
\end{figure*}

In the all-time averaged spectra, as shown in the top panel of \reffg{fig:process_rmbslsw},
there are still variations at the several mJy level. Such fluctuations may come from the sky background variation at different positions, or the change of system performance during observation, e.g. disparity of system temperature at different zenith angles \citep{2020RAA....20...64J}). These spectrally broad features can be separated from the narrow-frequency feature we are search for. To facilitate the subsequent processing, we have tried a number of approaches to fit the broadband spectral variation, such as median filtering, Hanning smoothing,  Extreme Envelope Curves (EEC) \citep{2022RAA....22h1001L}, and asymmetrically reweighted Penalized Least Squares (arPLS) \citep{baek2015baseline,2022arXiv220909555L}. In the end we chose the arPLS algorithm, which is excellent in getting an overall fitting curve while retaining the candidate signal. 

We take the data to be given by
\begin{equation} \label{eq:arpls_data}
    \mathbf{y = s + b + n}\,,
\end{equation}
where $\mathbf{s}$ denotes signal, $\mathbf{b}$ is the baseline, and $\mathbf{n}$ is the noise. The baseline $\mathbf{b}$ is determined by minimizing
\begin{equation} \label{eq:arpls_min}
    \mathbf{Q = (y-b)^TW(y-b) + \lambda b^TD^TDb\,, }
\end{equation}
where $\lambda$ is a smoothness parameter which is set to $10^3$ in this work, and $\mathbf{D}$ is a second order difference matrix, 
\begin{equation} \label{eq:arpls_D}
    \mathbf{D} = 
    \begin{bmatrix}
        1 & -2 & 1 & 0 & \cdots & 0 & 0 & 0 \\ 
        0 & 1 & -2 & 1 & \cdots & 0 & 0 & 0 \\
        \vdots & \vdots & \vdots & \vdots & \ddots & \vdots & \vdots & \vdots  \\
        0 & 0 & 0 & 0 & \cdots & 1 & -2 & 1
    \end{bmatrix}.
\end{equation}
The diagonal weight matrix $\mathbf{W}$ at the $i$th data point $y_i$ is defined as
\begin{equation} \label{eq:arpls_wi}
    w_i = 
    \begin{cases}
        \rm{logistic}(y_i-b_i,m_{d^-},\sigma_{d^-}),& \text{$y_i > b_i$}\\
        1,& \text{$y_i \le b_i$}
    \end{cases}
\end{equation}
where $m_{d^-}$ and $\sigma_{d^-}$ are the mean and standard deviation of $d^-$, which is in turn defined as the negative values of $\mathbf{d} \equiv \mathbf{y}-\mathbf{b}$. 
The logistic function is
\begin{equation}
    {\rm logistic}(d,m,\sigma) = \frac{1}{1+e^{k(d-(-m+s\sigma))}/\sigma}\,,
\end{equation}
where $k$ and $s$ are asymmetric and shifting coefficients which are usually set to be $k=2, s=2$ in arPLS. The initial $\mathbf{b}$ is calculated by applying an identity matrix $\mathbf{W}$, and the final $\mathbf{b}$ is obtained after several iterations until $\mathbf{W}$ becomes stable.
The parameter values are carefully chosen, to balance the fitness and smoothness. The residual baseline is then well subtracted as shown in the bottom panel of \reffg{fig:process_rmbslsw}. 

However, after performing the baseline subtraction, standing waves remain in the resulting spectrum. 
The reflection of radio waves between the reflector and feed cabin of FAST generate these standing waves, which show up as modulations with a period of $\sim1.1\,{\rm MHz}$ in the frequency spectrum. This has a large impact on the search for gegenschein lines, as the modulation period has a comparable spectral width. The amplitude and phase of the standing wave may vary over time. We tried a number of different techniques to remove this modulation in the data, including a model-dependent delay filter \citep{2009AJ....138..219P, 2021RAA....21...59L}, and model-independent methods such as component removal in Fourier space, which we ultimately use in our pipeline. 

\begin{figure}
    \centering
    \includegraphics[width=0.49\textwidth]{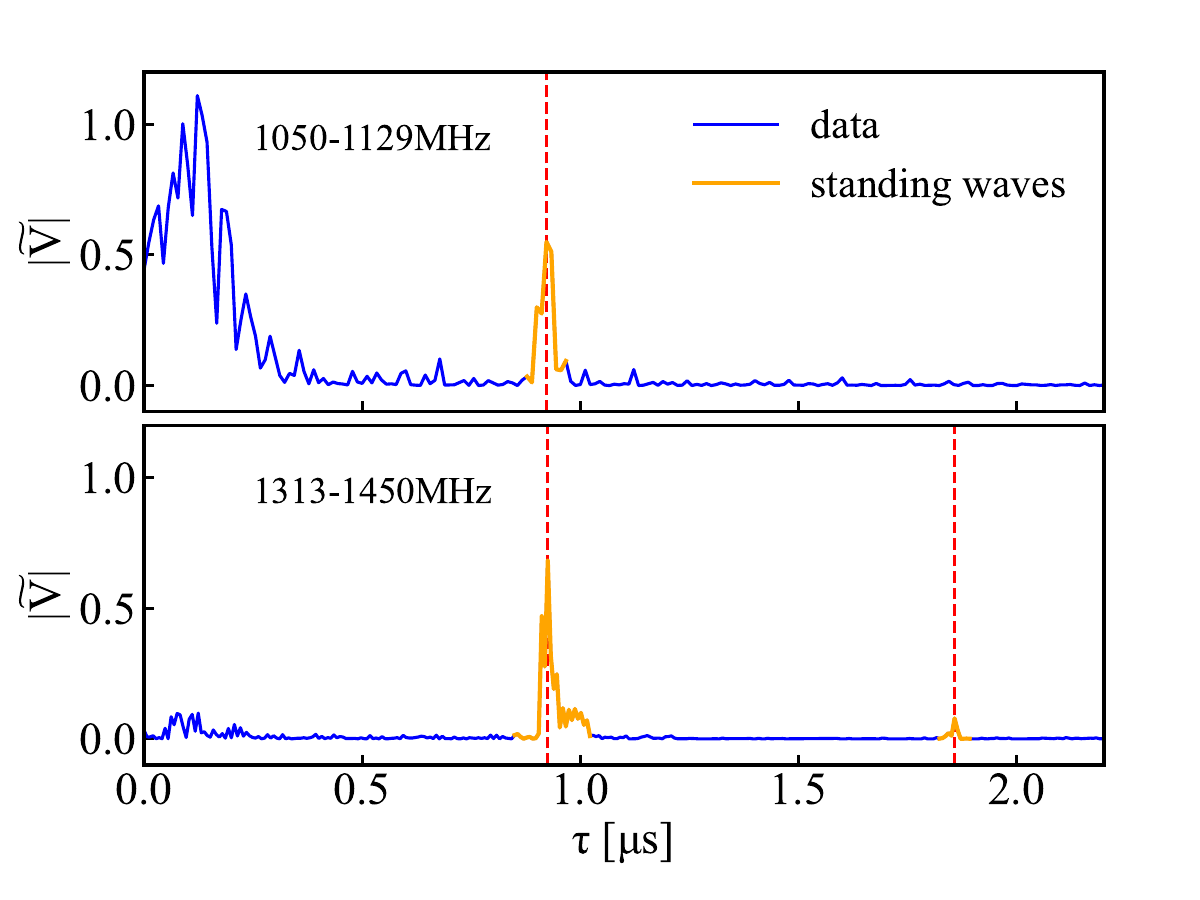}
    \caption{The delay spectra of the observed data in 1050-1135 MHz (upper panel) and 1313-1450 MHz (lower panel). The orange peaks are associated with the standing wave components that we need to remove, and the red dashed lines mark the delay $\tau$ corresponding to the ripples at frequency spectrum with period $\sim 1.1$ MHz (and its half frequency in the lower panel). }
    \label{fig:process_fft}
\end{figure} 

\begin{figure*}
    \centering
    \includegraphics[width=0.9\textwidth]{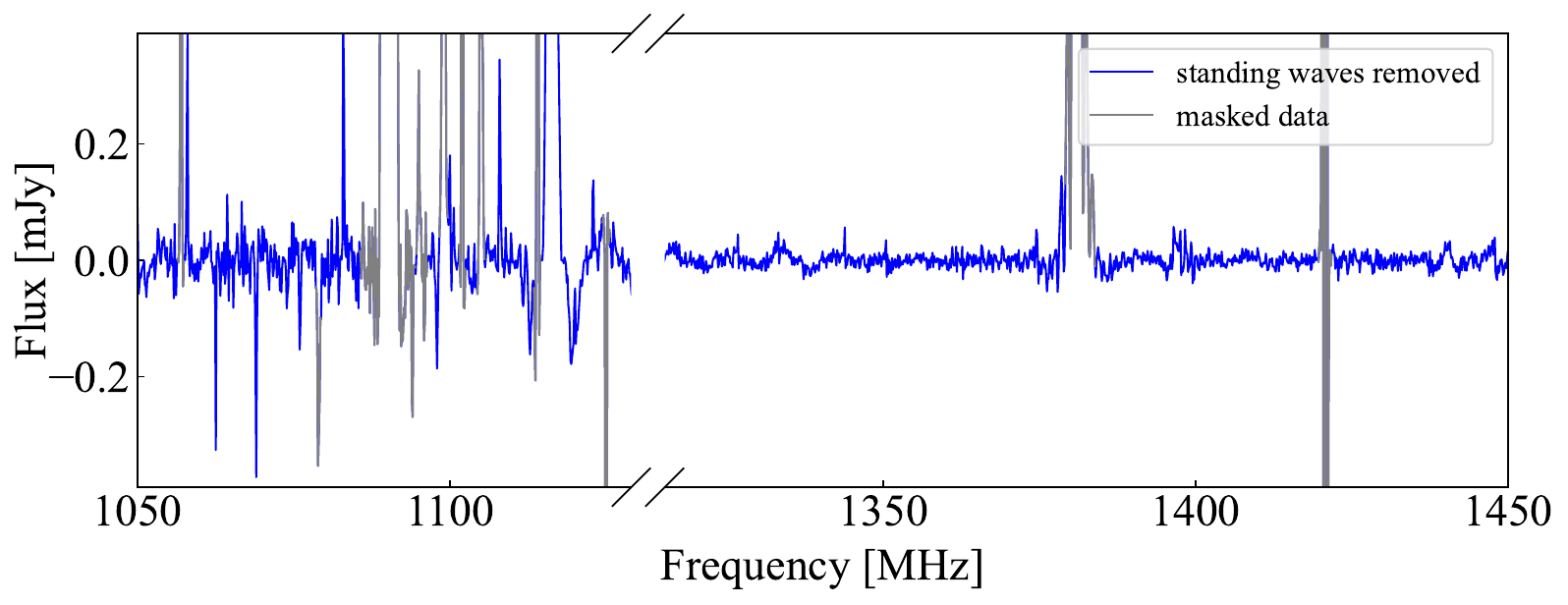}
    \caption{The final spectrum after baseline subtraction and standing waves removal at 1050-1129 MHz and 1313-1450 MHz. The grey lines show data which are severely contaminated by known strong RFIs and galactic HI signals.}
    \label{fig:process_result}
\end{figure*}

Before performing a delay transformation (i.e. Fourier transformation of the frequency spectrum), we mask frequency points which have been identified to be contaminated by known RFIs, as the RFIs are usually much stronger and would dominate the spectrum if left unmasked.  Instead of directly performing the Fourier transform, we use a Wiener Filter for delay spectrum estimation on the spectrum with masked gaps \citep{2022arXiv220201242C}.

In this approach, the frequency spectrum is modeled  as
\begin{equation}
    \mathbf{f = Fd + n}\,,
\end{equation}
where $\mathbf{f}$ is the frequency spectrum, $\mathbf{F}$ is the unitary Fourier transform matrix, $\mathbf{d}$ is the delay spectrum and $\mathbf{n}$ is the noise. The delay spectrum $\mathbf{d_w}$ is then estimated as
\begin{equation}
    \mathbf{d_w = (D^{-1} + F^{\dagger}N^{-1}F)^{-1} F^{\dagger}N^{-1}f }\,.
\end{equation}
where $\mathbf{N}$ is the noise covariance matrix with mask, $\mathbf{N^{-1}} = \sigma^{-2} \mathbf{M}$, where $\sigma$ is the root mean square (rms) of $\mathbf{f}$, and $\mathbf{M}$ is the mask matrix, with 0 at masked frequencies and 1 otherwise, $\mathbf{D}$ is the variance of the delay spectrum and is initially estimated as the variance of the Fourier transformation result of frequency spectrum. See the appendix of \citet{2022arXiv220201242C} for discussion on different delay spectrum estimators.

The delay spectra for our two working bands are shown in \reffg{fig:process_fft}.  The obvious peak at $\sim 9.09 ~\mu s$ shown in the orange line corresponds to the $\sim$ 1.1 MHz standing wave component. We extract the standing wave component by performing an inverse Fourier transform of the peaks in the delay spectrum. 

We have checked whether this procedure would remove the signal we are seeking by adding a mock signal in the spectrum, and find that it will not. We then perform the baseline subtraction again using a median filter with a window width $\sim 6 \MHz$ for baseline fitting, to remove small fluctuations which have been hidden in the standing waves before, and further mask frequency points which contain $> 20\%$ refilled values as mentioned in \refsc{subsec:process_RFI}. 
After subtracting these components, we obtain the result shown in \reffg{fig:process_result}. 

To improve the final signal-to-noise ratio of the spectrum, we also perform a weighted average of the data for all days, beams, and polarizations except for the bad data. The weighting factor for the data of each day is set to be proportional to the total integration time of that day, while the weighting factor for different beams are set to be proportional to the expected flux density at that beam. The data from the beams near the expected gegenschein image are given higher weights, in order to maximize the signal-to-noise ratio. 

As noted above, the position of the Vela gegenschein is near the galactic plane, where the sky background has larger spatial variations. We have observed four different OFF-source positions (shown as white circles in \reffg{fig:src_beam}) adjacent to the Vela antipodal point to reduce the influence of sky background variation. After the baseline subtraction by applying the arPLS method and standing wave removal, the residual ON-OFF data for all OFF-source positions are expected to yield the same gegenschein signal $S_0$, i.e. 
$\rm  S_{ON-OFF1} = S_{ON-OFF2} = S_{ON-OFF3} = S_{ON-OFF4} \equiv S_0$. 
We can then construct an estimator of $S_0$ as the weighted sum of these. 

If the noise is pure thermal noise, with ${\rm \sigma^2_{ON-OFF_i}} \propto 1/t_i$, where $t_i$ is the effective integral time of the $i$th ON and OFF observation as in \reftb{tb:obs}, the weights to maximize the signal to noise ratio  
should be $$w_{{\rm OFF},i}=\frac{t_i}{t_{\rm total}},$$ where $i=1,2,3,4$, and $t_{\rm total}=\sum_i t_i$, which would yield $w_{\rm OFF,1}=0.473,~ w_{\rm OFF,2}=0.164, ~w_{\rm OFF,3}=0.182,$ and $w_{\rm OFF,4}=0.182$. 

However, in reality the noise contains significant non-thermal components, such as residual
sky background, RFI contamination, baseline, etc., so the total noise does not follow the thermal noise relation $\sigma_i \propto t_i^{1/2}$.  More importantly, the signals in the ON-OFF data spectrum are not all the same, because of signal loss while subtracting different OFF-source data. 
Therefore, we use an alternative strategy to obtain the best weights. We leave $w_{{\rm OFF},i}$ as a free parameter in the range $(0,1)$, with the constraint $\sum_i w_{{\rm OFF},i}=1$, and vary them to maximize 
$${\rm S/N} = \frac{1}{\sigma_{\rm tot}}\sum_i (\omega_{{\rm OFF}_i} \cdot S_{{\rm ON-OFF}_i}).$$ 
Note that $S_{{\rm ON-OFF}_i}$ are calculated using the beam model and the gegenschein model (Eq.~\ref{eq:S_obs}) normalized to eliminate the %uncertain 
unknown signal parameter $g_{a\gamma\gamma}$, and the noise $N$ is estimated with our weighted averaged spectrum given $w_{\rm OFF i}$. 
In this way, the weights of the four OFF-source points are found to be: 
\begin{itemize}
    \item 1050-1129 MHz:\\
    $w_{\rm OFF1} = 0.115,\, w_{\rm OFF2} = 0.045, \\ w_{\rm OFF3} = 0.563, w_{\rm OFF4} = 0.277$; 
    \item 1313-1450 MHz:\\ 
    $w_{\rm OFF1} = 0.413,\, w_{\rm OFF2} = 0.117, \\ w_{\rm OFF3} = 0.122, w_{\rm OFF4} = 0.348$.
\end{itemize}

\reffg{fig:process_result} is the final spectrum with these weights. After processing the data, the final spectrum is ready for performing a search for the target signal. 
The residual spikes in \reffg{fig:process_result} are caused either by RFIs, which include those which are stable all the time or too weak to be flagged, or by fluctuations with width similar to our target signal. To avoid removing target signal, these residual spikes are kept at this stage.

\section{Results} \label{sec:result}

\subsection{Detection limit} \label{subsec:result_limit}

We first estimate the sensitivity of FAST with parameters in this observation using the general radiometer equation 
\begin{equation}\label{eq:sensitivity}
\sigma_{\rm N} = \frac{2 k_{\rm B} 
T_{\rm sys}}{\eta_{\rm A} A_{\rm illu}
\sqrt{N_{\rm pol}\Delta \nu \Delta t}}.
\end{equation}
The system temperature $T_{\rm sys} \sim 25\ {\rm K}$, aperture efficiency $\eta_{\rm A} \sim 0.7$, and the illuminated area of FAST with a diameter of 300m has $A_{\rm illu} \sim 70700\ {\rm m}^2$, and $N_{\rm pol} = 2$ is the number of polarization channels. We take the frequency resolution to be $\Delta \nu = 0.12\ {\rm MHz}$. 
The integration time $\Delta t$ adopted here varies on different days. For data set A, the time allocation is $t_{\rm ON} = 120\,{\rm minutes}, t_{\rm OFF} = 10\,{\rm minutes}$. 
For data set B, for the first two days, $t_{\rm ON} = t_{\rm OFF} = 70\,{\rm minutes}$. For the last three days, $t_{\rm ON} = t_{\rm OFF} = 60\,{\rm minutes}$. For data set C, there are 4 days of observation, and $t_{\rm ON} = t_{\rm OFF} = 72\,{\rm minutes}$ every day. 

The combined noise level for each beam in each day should be 
$$\sigma_{\rm N,beam,day} = \sqrt{\sigma_{\rm N, ON,beam,day}^2 + \sigma_{\rm N, OFF,beam,day}^2},$$  
which is approximately equivalent to the observation duration of $\sim$ 793 minutes ON-source tracking and $\sim$ 793 minutes OFF-source tracking in total. After that, we consider the weights for different beams and OFF-source points as described in \refsc{subsec:process_rmbslsw} before stacking them together and calculating the total noise level, i.e. 
\begin{equation}
    \sigma_{\rm N, total} = \sqrt{\sum_{\rm day=1}^{19}(\omega_{\rm day}^2 \sum_{\rm beam=1}^{19}(\omega_{\rm beam}\sigma_{\rm N,beam,day})^2)}\,.
\end{equation}
The estimation gives expected values of $\sim 6.3\ \mu\mathrm{Jy}$ at 1050-1129 MHz and $\sim 4.7\ \mu\mathrm{Jy}$ at 1313-1450 MHz.

In the actual observation data, the RMS of final spectrum is $\sim 36.4 \mu\mathrm{Jy}$ for the 1050-1129 MHz band, and $\sim 13.0 \mu\mathrm{Jy}$ for the 1313-1450 MHz band, which are  $\sim$ 5.8 and $\sim$ 2.8 times higher than the expected noise, respectively. The higher-than-expected noise level indicates that the noise in the final spectrum is not purely thermal noise. 

To derive an upper limit on $g_{a\gamma\gamma}$ from  our observations, we use a method similar to \citet{2022PhRvD.106h3006Z} to calculate the probability distribution of $g_{a\gamma\gamma}$ at each frequency channel. The likelihood function is 
\begin{equation} \label{eq:llh}
    \mathcal{L}(m_a,g_{a\gamma\gamma}) = \prod \limits_{i}^{\rm N_{ch}} \frac{1}{\sqrt{2\pi}\sigma_{\nu_i}} \mathrm{exp}[-\frac{d_{\nu_i}-\mu_{\nu_i}-\overline{S}_{\nu_i}(m_a,g_{a\gamma\gamma})^2}{2\sigma_{\nu_i}^2}] \,,
\end{equation}
where $d_{v_i}$ is the data at frequency $\nu_i$, $\mu_{\nu_i}$ is the mean value of the background at $\nu_i$, which is set to zero here as we have already subtracted the continuum baseline, $\overline{S}_{\nu_i}$ is the theoretical flux spectrum value at $\nu_i$ with given axion mass $m_a$ and $g_{a\gamma\gamma}$, and $\sigma_{\nu_i}$ is the standard deviation of the noise at that frequency, as estimated from the data in the nearby $\sim$ 10 MHz and corrected by the correction factor $C_{\rm mask}(\nu_i) = \sqrt{1-f_{\rm mask}(\nu_i)}$ where $f_{\rm mask}(\nu_i)$ is the mask fraction as introduced in \refsc{subsec:process_RFI}. We assume that the gegenschein emission signal has a Gaussian frequency profile, centered  at $m_a$ with the width $\sigma_i$, given by 
$\sigma_i = \nu_i \sigma_d / c$, with the dark matter velocity dispersion $\sigma_d = 116\mathrm{km/s}$. 
Given the theoretical gegenschein signal model and FAST beam pattern, the relation between the peak flux $S_{\nu_i}$ and the axion-photon coupling strength $g_{a\gamma\gamma}$ can be calculated with the integral flux,
\begin{equation} \label{eq:S_theo}
     \overline{S}_{\mathrm{obs}}(g_{a\gamma\gamma}) = \frac{\hbar c^4 g_{a\gamma\gamma}^2}{16}\iint I_{\nu}(x_d,\hat{n}) \rho(x_d) b(\hat{n}) dx_d d\Omega \,.
\end{equation}
For example, for a mock gegenschein signal centered at 1400 MHz,
\begin{equation}
    S_\nu(\nu_c=1400\MHz) = 2.2  \left(\frac{g_{a\gamma\gamma}}{10^{-10}\mathrm{GeV}^{-1}}\right)^2\,\mu\mathrm{Jy}\,.
\end{equation}
This value varies with the different central frequency of the axion signal.

\begin{figure}
    \centering
    \includegraphics[width=0.47\textwidth]{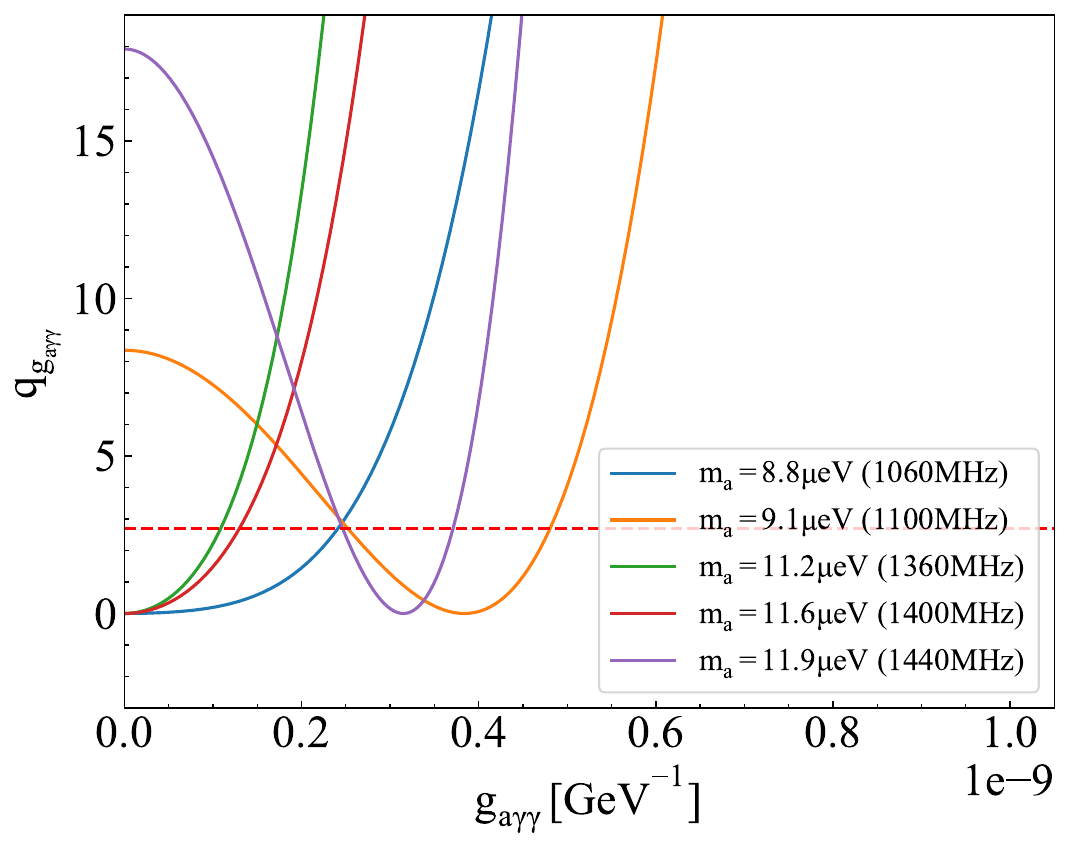}
    \caption{The test statistic $-2{\rm ln}\,\Lambda (g_{a\gamma\gamma})$ at 5 different values of the axion mass as examples. The horizontal dashed line is $q_{g_{a\gamma\gamma}} = 2.71$, with regions to the right of the intersection being ruled out at 95\% confidence for each value of $m_a$. 
   }
    \label{fig:logllh}
\end{figure}

Following the analysis in \cite{2011EPJC...71.1554C}, we use the likelihood ratio $\Lambda (g_{a\gamma\gamma})$ and corresponding upper-limit test statistic to constrain $g_{a\gamma\gamma}$. 
The test statistic $q_{g_{a\gamma\gamma}}$
at fixed $m_a$ and $g_{a\gamma\gamma}$ is defined as 
\begin{equation}\label{eq:llr}
    q_{g_{a\gamma\gamma}} = -2{\rm ln}\Lambda (g_{a\gamma\gamma}) = -2 {\rm ln} \frac{\mathcal{L}(m_a,g_{a\gamma\gamma})}{\mathcal{L}_{\rm max}}\,,
\end{equation}
where $\mathcal{L}_{\rm max}= \mathcal{L}(m_a, \hat{g}_{a\gamma\gamma}$ is the maximum likelihood, and $\hat{g}_{a\gamma\gamma}$ is the maximum likelihood estimator, i.e. $\hat{g}_{a\gamma\gamma} = {\rm arg\, max}_{g_{a\gamma\gamma}} \mathcal{L}(m_a,g_{a\gamma\gamma})$. 
This test statistic is only meaningful for $g_{a\gamma\gamma}>\hat g_{a\gamma\gamma}$ when the likelihood tests have enough constraining power \citep{2011arXiv1105.3166C}. 
Examples of this test statistic at five different values of $m_a$ (corresponding to five frequency channels) are shown in \reffg{fig:logllh}. 
The value of $q_{g_{a\gamma\gamma}}(g_{a\gamma\gamma}=C)$ represents the strength of evidence to exclude the parameter space $g_{a\gamma\gamma} > C$, where $C$ is a constant satisfying $C>\hat{g}_{a\gamma\gamma}$. The value $g_{a\gamma\gamma} = \hat{g}_{a\gamma\gamma}$ corresponds to the minimum value of the upper-limit test statistic, indicating the maximum-likelihood value of $g_{a\gamma\gamma}$.
For a pure noise spectrum without real or spurious signals (i.e. under the null hypothesis), the minimum point of $-2{\rm ln}\Lambda (g_{a\gamma\gamma})$ is expected to be at $g_{a\gamma\gamma} = 0$, as the blue, green and red curves in \reffg{fig:logllh}. 
In contrast, the curves with $\hat{g}_{a\gamma\gamma} > 0$ indicate an excess in this frequency channel that may be a potential axion echo signal, as the orange and purple curves in \reffg{fig:logllh}. In \refsc{subsec:result_signal}, we quantify the significance of each candidate signal and perform detailed checks to rule out the signal hypothesis.

Since the upper-limit test statistic 
aligns with the asymptotic $\chi^2$ distribution for one degree of freedom \citep{10.1214/aoms/1177732360}, the 95\% upper limit of $g_{a\gamma\gamma}$ is given by $-2{\rm ln}\Lambda (g_{a\gamma\gamma}) \leq 2.71$ \citep{2011EPJC...71.1554C,2011arXiv1105.3166C}, which is the intersection of the colored curves and the horizontal red dashed line at $g_{a\gamma\gamma} > \hat{g}_{a\gamma\gamma}$ in \reffg{fig:logllh}. By constructing the upper-limit test statistic for each frequency channel, we can obtain the 95\% upper limit on $g_{a\gamma\gamma}$ at each corresponding $m_a$.

\begin{figure*}
    \centering
    \includegraphics[width=0.32\textwidth]{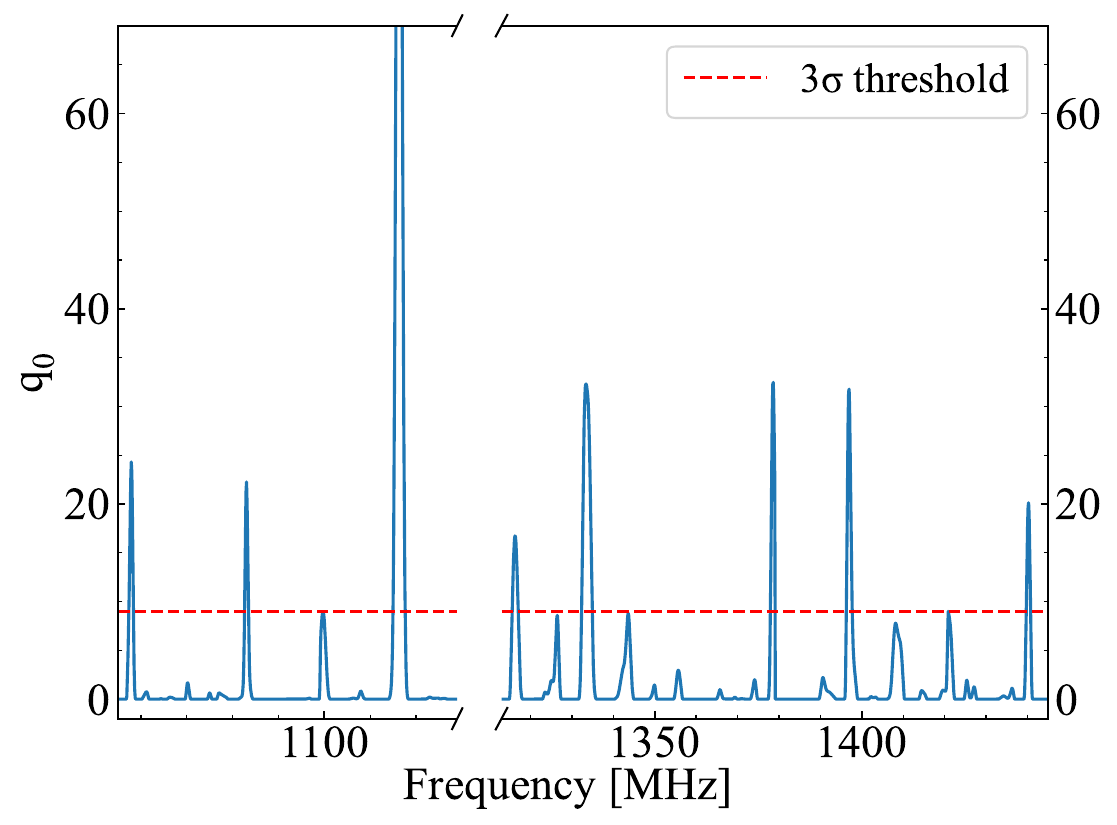}
    \includegraphics[width=0.325\textwidth]{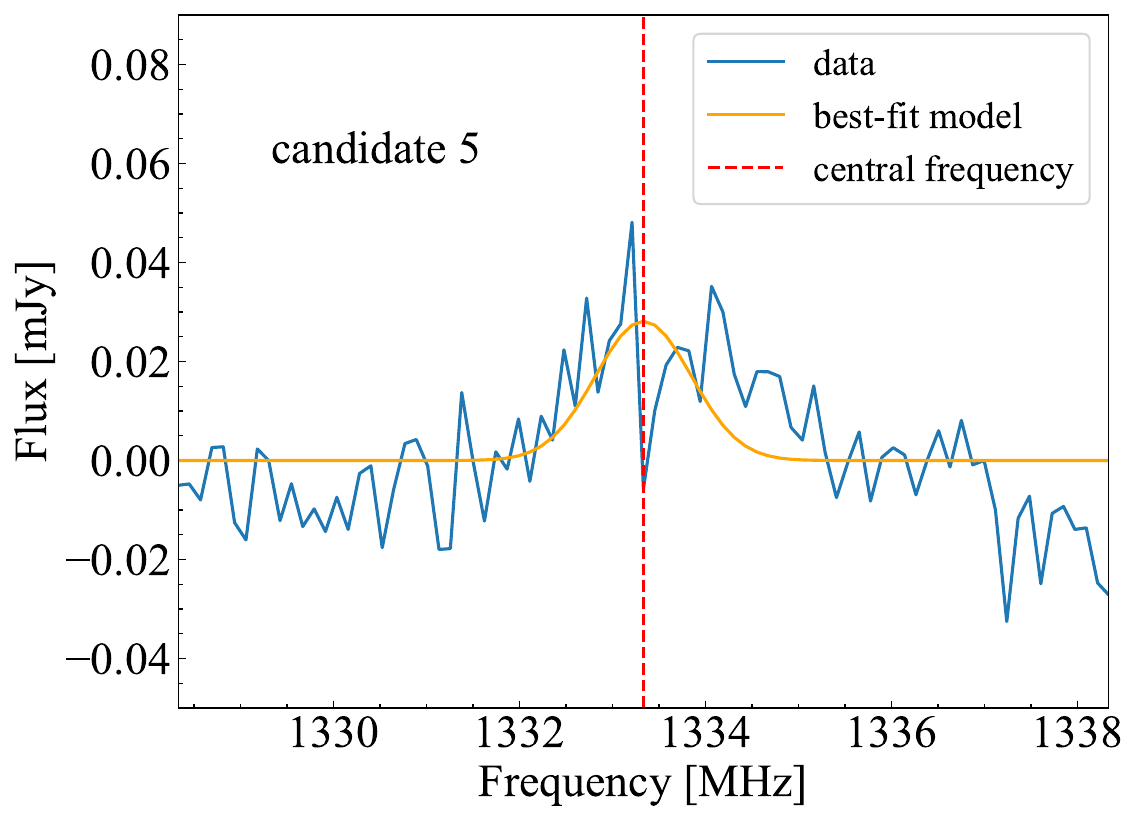}
    \includegraphics[width=0.32\textwidth]{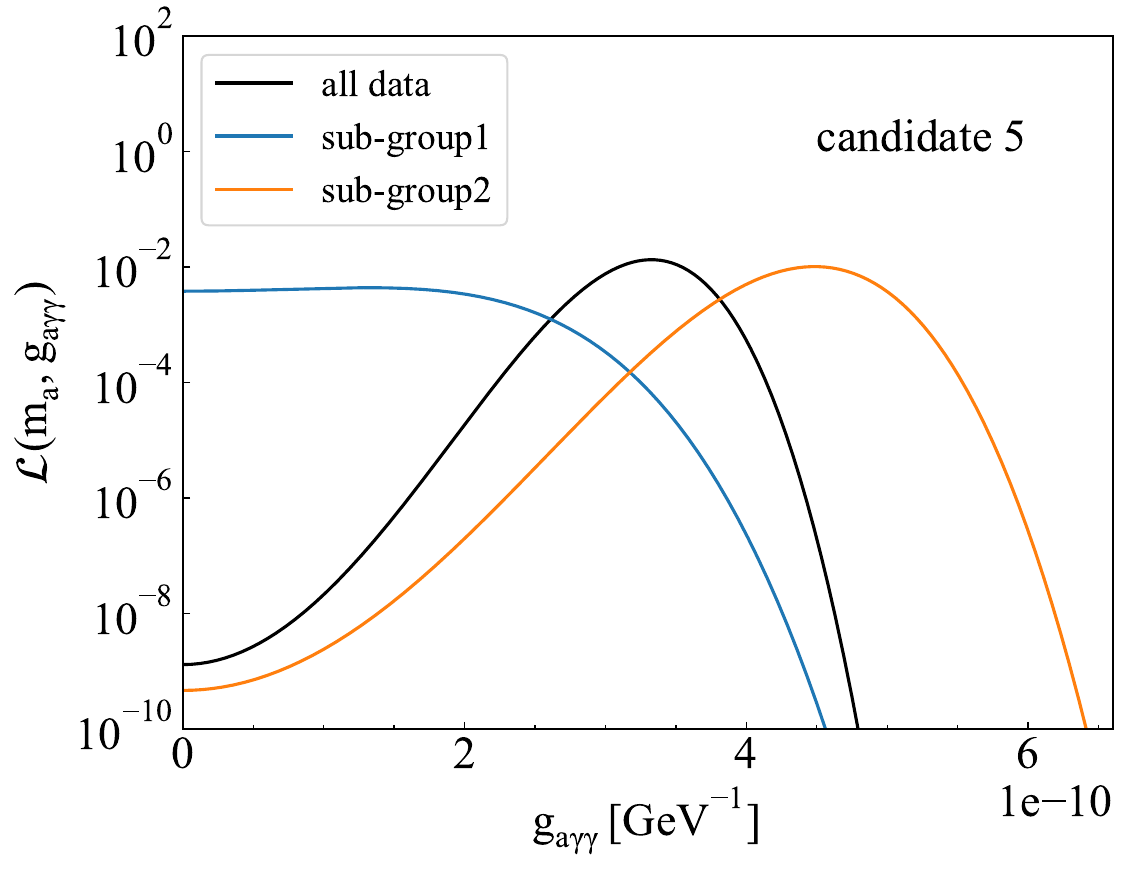}
    \caption{Left: the blue line shows the $q_0$ at different frequency channels. Middle: the spectrum of candidate 5 as an example. The blue line is the observed spectrum, the orange is the best-fit model and the red dashed line marks the central frequency. Right: the likelihood distribution 
    of candidate 5 with the spectra of two sub-groups at the frequency channel $\nu_c$ as an example.
   }
    \label{fig:TS}
\end{figure*}

The constraints derived from our observations are shown as the blue solid line in \reffg{fig:detect_limit}. The gaps at frequency $\sim$ 1090 MHz and $\sim$ 1380 MHz coincide with RFI-flagged regions. The constraint on $g_{a\gamma\gamma}$ reaches $\sim 2 \times 10^{-10}\ \rm GeV^{-1}$, which is slightly weaker than the theoretical limit at most frequency channels, as shown by the black dashed line which is derived from mock data with pure thermal noise assuming the same observation parameters. We discuss the possible reasons for these excesses in \refsc{sec:discuss}.  Our results provide a stronger constraint on $g_{a\gamma\gamma}$ in this axion mass range compared with results obtained by a direct search for a possible axion decay signal from galaxy clusters which uses FAST observations \citep{2024PhLB..85238631G}, but is a factor of $\sim 3$ times weaker than the CAST limit $g_{a\gamma\gamma} \le 6.6 \times 10^{-11}{\rm GeV^{-1}}$ \citep{2017NatPh..13..584A}.

\subsection{Candidate signal} \label{subsec:result_signal}

With some excesses seen in the likelihood ratios, we have systematically checked the possibility of detecting axion gegenschein at the frequencies where excesses are observed.
Our procedures are as follows: 
\begin{enumerate}
    \item \textbf{Primary search:} We calculate the discovery test statistic $q_0$ at fixed $m_a$ similar to \refeq{eq:llr} to quantify the significance of a positive signal \citep{2011EPJC...71.1554C}.
    The discovery test statistic $q_0$ here is analogously defined as twice the logarithm likelihood difference between the best-fit model and the null model which has $g_{a\gamma\gamma} = 0$, i.e. 
    \begin{align}\label{eq:llr0}
        q_0
        &= -2\,{\rm ln} \frac{\mathcal{L}(m_a,g_{a\gamma\gamma}=0)}{\mathcal{L}_{\rm max}}
        %(m_a,\hat{g}_{a\gamma\gamma})}\,. 
        % \notag \\
        % &= {\rm max}_{g_{a\gamma\gamma}} 2\, {\rm ln}\Lambda_0 (g_{a\gamma\gamma})\,,
    \end{align}
    The $q_0$ values at all frequency channels are shown in the left plot of \reffg{fig:TS}. We record frequency channels with $q_0 > 9$ corresponding to the 3-sigma detection \citep{2011EPJC...71.1554C} 
    as potential signal locations, where the probability of a signal located at $\hat{g}_{a\gamma\gamma}$ is significantly larger than the null case $g_{a\gamma\gamma} = 0$. 
    \item \textbf{Candidate combination:} If potential signals appears at a number of adjacent channels between $[\nu_1, \nu_2]$, we combine them as one candidate because the expected signal profile with standard deviation $\sigma_i \approx 0.5\MHz$ is larger than the frequency resolution of our data ($\Delta \nu = 0.12 \MHz$). We have also checked and found that the width of each combined candidate is not larger than the expected signal width. 
    We record the frequency channel with the largest $q_0$ across $[\nu_1, \nu_2]$ as the central frequency $\nu_c$ of this candidate. The spectrum of one candidate (near 1333.4 MHz) is shown in the middle plot of \reffg{fig:TS} as an example. We can see an excess at the marked frequency channel. 
    \item \textbf{Cross-check:} To check if the candidates are real signals or false identifications, we divide the data into different sub-groups according to the observation time and cross-check between them. Specifically, we choose two sub-groups with the same integration time $\Delta t \approx 270$ minutes and theoretical signal loss. If one candidate is a real axion gegenschein signal, we could expect the consistency of detection between different sub-groups and the total data because the signal is supposed to be stable with time. We calculate the likelihood for the spectra of each group based on \refeq{eq:llh}. To quantify the consistency, we calculate the $q_{g_{a\gamma\gamma},i}$ as the likelihood statistic of the $i$-th sub-group ($i=1,2$) by
    \begin{align}\label{eq:llr_subgroup}
        q_{g_{a\gamma\gamma},i} 
        &= -2\,{\rm ln} \frac{\mathcal{L}_i(m_a,g_{a\gamma\gamma,i}=\hat{g}_{a\gamma\gamma})}{\mathcal{L}_i(m_a,\hat{g}_{a\gamma\gamma,i})}\,,
    \end{align}
    where $\mathcal{L}_i(m_a, g_{a\gamma\gamma,i})$ is the likelihood function with data in the $i$-th sub-group, $\hat{g}_{a\gamma\gamma}$ and $\hat{g}_{a\gamma\gamma,i}$ are the maximum likelihood estimator for total data and the $i$-th sub-group data, respectively. An example of the $\mathcal{L}(m_a,g_{a\gamma\gamma,i})$ for one candidate is shown in the right plot of \reffg{fig:TS}. If $q_{g_{a\gamma\gamma},i} > 2.71$ 
    for either sub-group, we would regard this candidate as a misidentification at 95\% confidence level. In other words if $\hat{g}_{a\gamma\gamma}$ is ruled out by a subset of the data at more than 95\% confidence, we regard any excess in the overall data as spurious.
\end{enumerate}

Using our observational data and following this procedure, we find a total of 8 candidates after the first two steps. However, none of them passed the cross-check. Thus, in the present search we do not make a positive detection and can only derive upper limits on $g_{a\gamma\gamma}$. With a more refined understanding of the signal characteristics, further accumulation of observational data, and continued optimization of our data processing techniques, we aim to enhance our ability to identify such signals in the future.

\section{Discussion} \label{sec:discuss}

As can be seen in \reffg{fig:detect_limit}, the actual sensitivity as estimated from the data is a few times lower than the theoretical limit, showing that there are sources of error aside from the thermal noise. In this section, we discuss the potential origins of the excess error, as well as issues which may affect our results.

\subsection{Calibration error} \label{subsec:discuss_cal}
The calibration process is itself affected by thermal noise. Taking data B as an example and using $\rm V_{on}-V_{off}$ to obtain the bandpass, for $\rm V_{on}$ the integral time in 10 minutes is 75~s, and for $\rm V_{off}$ it is 525~s, the estimated thermal noise is 8.26~mK in $\rm V_{on}$, and 3.12~mK in $\rm V_{off}$ for each 0.122 MHz frequency bin, the estimated total thermal noise is then 8.84 mK. Considering that $\rm V_{on}-V_{off}$ corresponds to $\rm T_{on}-T_{off} = T_{ND} \sim 1K$, the relative bandpass error caused by thermal noise would be at a level of $\rm 8.84mK/1K \sim 0.9\%$, consistent with the above estimates.

Now we look at the real data and examine how large is the residue error. 
In \reffg{fig:bpnorm} we show the change in bandpass shape for each 10-minute time block. As the figure shows, the variation is at the $\lesssim 1\%$ level for one day of observation (one beam and one polarization).

\begin{figure}
    \centering
    \includegraphics[width=0.47\textwidth]{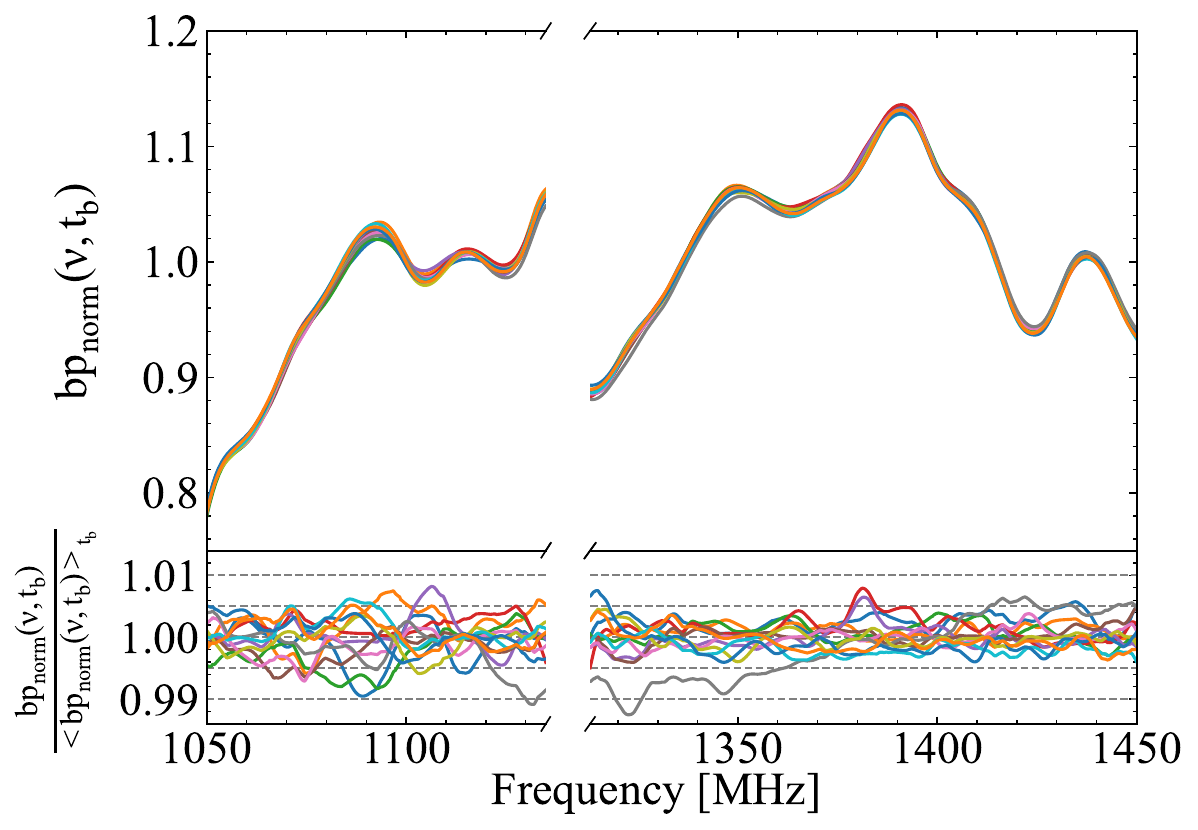}
    \caption{Comparison of normalized bandpass (normalized to data B, Day 5, M01, xx polarization). Lines in the upper panel show bandpass normalized by the median value, while lines in the lower panel display the relative bandpass shape in comparison with the average of all time blocks. Lines with different color represents different time blocks. 
}
    \label{fig:bpnorm}
\end{figure}

In the middle panel of \reffg{fig:compare_ONOFF}, the ON-OFF spectra on different days are shown, and they vary significantly. This may indicate that there is larger error in the bandpass from the ON-OFF measurement than expected from thermal noise. 
Concretely, if we denote the sky background intensity at the ON position as $b_0$, at the OFF$_i$ position as $b_i$, and the gegenschein signal we are searching as $s$, then the ON-OFF$_i$ difference for day $\alpha$ is

\begin{align}
\Delta_{i,\alpha}&= V_{{\rm ON},\alpha} - V_{{\rm OFF_i},\alpha} \notag \\
&=g_{\alpha} (b_0 + s) - (g_{\alpha}+\delta g_{\alpha})b_i  \notag \\
&=g_{\alpha}(b_0-b_i) - \delta g_{\alpha} b_i + g_{\alpha} s, 
\label{eq:diff_i2}
\end{align}
where $g_\alpha$ is the average gain on day $\alpha$ when observing the ON region, and $\delta g_{\alpha}$ is the average gain difference when observing the ON and OFF$_i$ region on day $\alpha$, i.e. the variation within the day. In this expression we have neglected radio interference and thermal noise. 
The signal $s$ should be narrow in frequency, so the ON-OFF$_i$ differences over wide frequency range should be due to the first and second terms in the last line of Eq.~(\eqref{eq:diff_i2}), while the day-to-day variation should be due to variation in $g_{\alpha}$ and $\delta g_{\alpha}$.

From the left panel of Fig.~\ref{fig:compare_ONOFF} which shows a one-day observation result, we can see that on this day, at the 1050-1150 band, the flux $g_\alpha b_3 \sim 2.2$ Jy, while at the 1300-1400 band, $g_\alpha b_3 \sim 1.8$ Jy, so for a rough estimate, we can take $g_\alpha b_3 \sim 2$ Jy. Also from this figure we can see that the first term in Eq.~(\eqref{eq:diff_i2}), $g_{\alpha}(b_0-b_3) \sim 0.1 \Jy$, thus the second term, using the $g_\alpha b_3 \sim 2$ Jy derived above, is at the $\sim 2 \frac{\delta g_\alpha}{g_\alpha}$ Jy level.

With these estimates, we can look at the day-to-day variation of $\Delta_3={\rm ON}-{\rm OFF}_3$ shown in the middle panel of \reffg{fig:compare_ONOFF} again. For different days $\alpha$, the amplitude $\Delta_3$ varies between 0.01 and 0.1 Jy, which is comparable with $\Delta_3$ itself. If this is due mainly to the first term in \refeq{eq:diff_i2}, then it arises from the variation of $g_\alpha$, but as the gain $g_\alpha$ is absolutely calibrated, it is unlikely that there is such large variation. More likely, it originates from the second term in \refeq{eq:diff_i2}, due to the variation in $\delta g_\alpha$. Taking the total range of variation as $0.1$ Jy, and using the estimate of the second term obtained in the last paragraph $\sim 2 \frac{\delta g_\alpha}{g_\alpha}$ Jy, we estimate $\frac{\delta g_\alpha}{g_\alpha} \sim 5\%$. However, the value in this estimation is more like an upper limit in the worst case, while the typical variation is at the 1\% level. %Therefore, we conservatively regard it at the $1\%$ level. 

Finally, the right panel of \reffg{fig:compare_ONOFF} shows the result for subtracting different blank field. Each blank field may have different background intensity and spectrum, so some variation is expected. These results are derived for the combination of several days observation, so the magnitude is smaller than the one day observations shown in the left and middle panels. The ${\rm ON-OFF_2}$ and  ${\rm ON-OFF_3}$ spectra are similar, while  ${\rm ON-OFF_1}$ and  ${\rm ON-OFF_4}$ spectra are similar, this is likely due to the similarity between the OFF$_2$ and OFF$_3$ spectra, as well as between OFF$_1$ and OFF$_4$ spectra. 

We also compare the bandpass error of the two different calibration modes for data sets A and B as described in \refsc{sec:obs}. We define the residual normalized bandpass $g_{\rm err}$ as
\begin{equation} \label{eq:gerr}
    g_{\rm err}(\nu, t) = g_{\rm norm}(\nu, t) - [g(\nu)g(t)]_{\rm norm} \,
\end{equation}
to represent the error between the real bandpass $g_{\rm norm}(\nu, t)$ and our estimation $[g(\nu)g(t)]_{\rm norm}$, where $g_{\rm norm}(\nu, t)$ is given by the unsmoothed $V_{\rm on}(\nu, t) - V_{\rm off}(\nu, t)$ normalized by their median value, $g(\nu)$ is the smoothed bandpass variation, 
% \ks{?} \wy{Revised}
and $g(t)$ is the smoothed temporal fluctuation of the bandpass. Here the time resolution is set to be $\sim$8~s (the same as the noise injection period in data set B) for easier comparison. The histograms of $g_{err}$ and the Gaussian fits are shown in \reffg{fig:dbp_hist}. The overall deviation estimated from the mean values of $g_{err}$ are $\mu_A = -0.002,\, \mu_B = 0.010$, indicating an error of $\lesssim 1\%$ during our bandpass estimation process. The standard deviation of the Gaussian component for data set A is slightly smaller than data set B. There are non-gaussian tails in the histogram representing large $g_{err}$, but they are still a very small fraction ($\rm N/N_{total} \lesssim 10^{-4}$) of the total, and are similar for the two data sets. 

\begin{figure}
    \centering
    \includegraphics[width=0.45\textwidth]{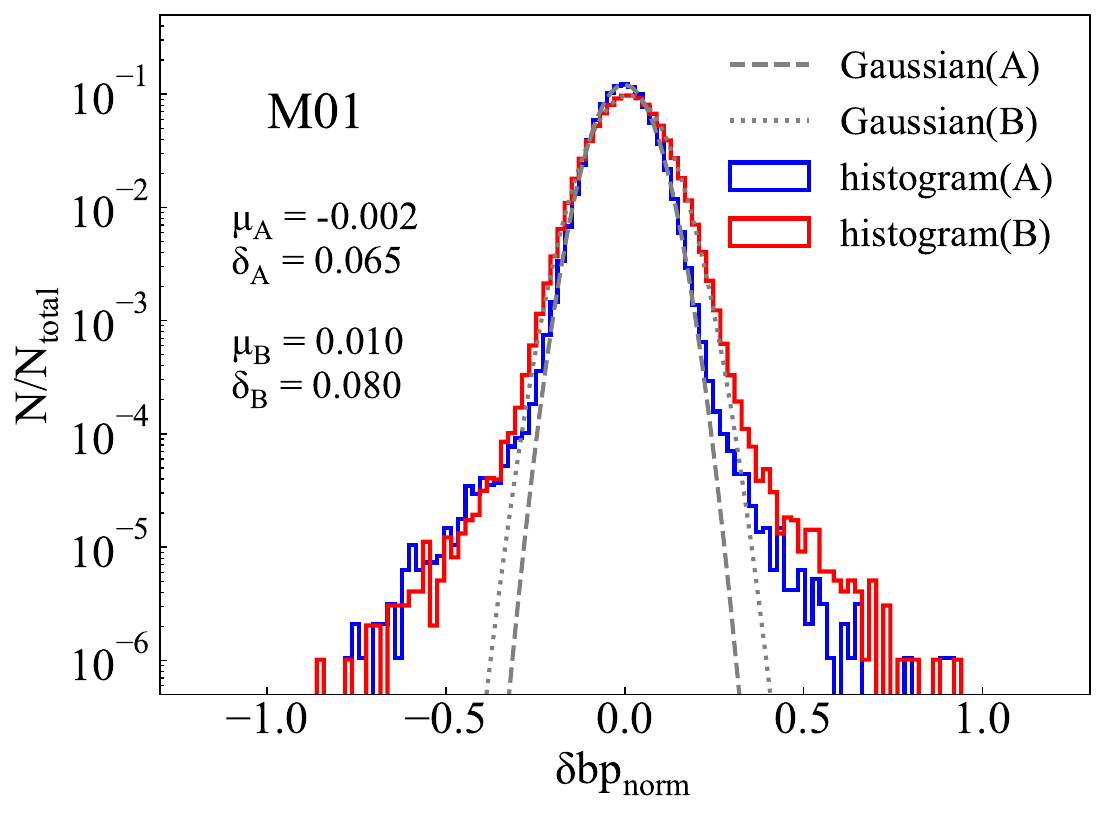}
    \caption{Histograms of the residual normalized bandpass $g_{\rm err}(\nu, t)$ in Day 1 of data set A and Day 5 of data set B. Statistical results of data set A and B are shown by red curve and the blue curve respectively. The Gaussian fitted results of the histograms are plotted with grey dashed and dotted line. }
    \label{fig:dbp_hist}
\end{figure}

Based on the analysis above, we estimate the flux error due to bandpass calibration as $\sim 1\%$. Its influence for the constraint of $g_{a\gamma\gamma}$ is shown in \reffg{fig:calibration_err}, which is $\sim 0.5\%$.

\begin{figure}
    \centering
    \includegraphics[width=0.49\textwidth]{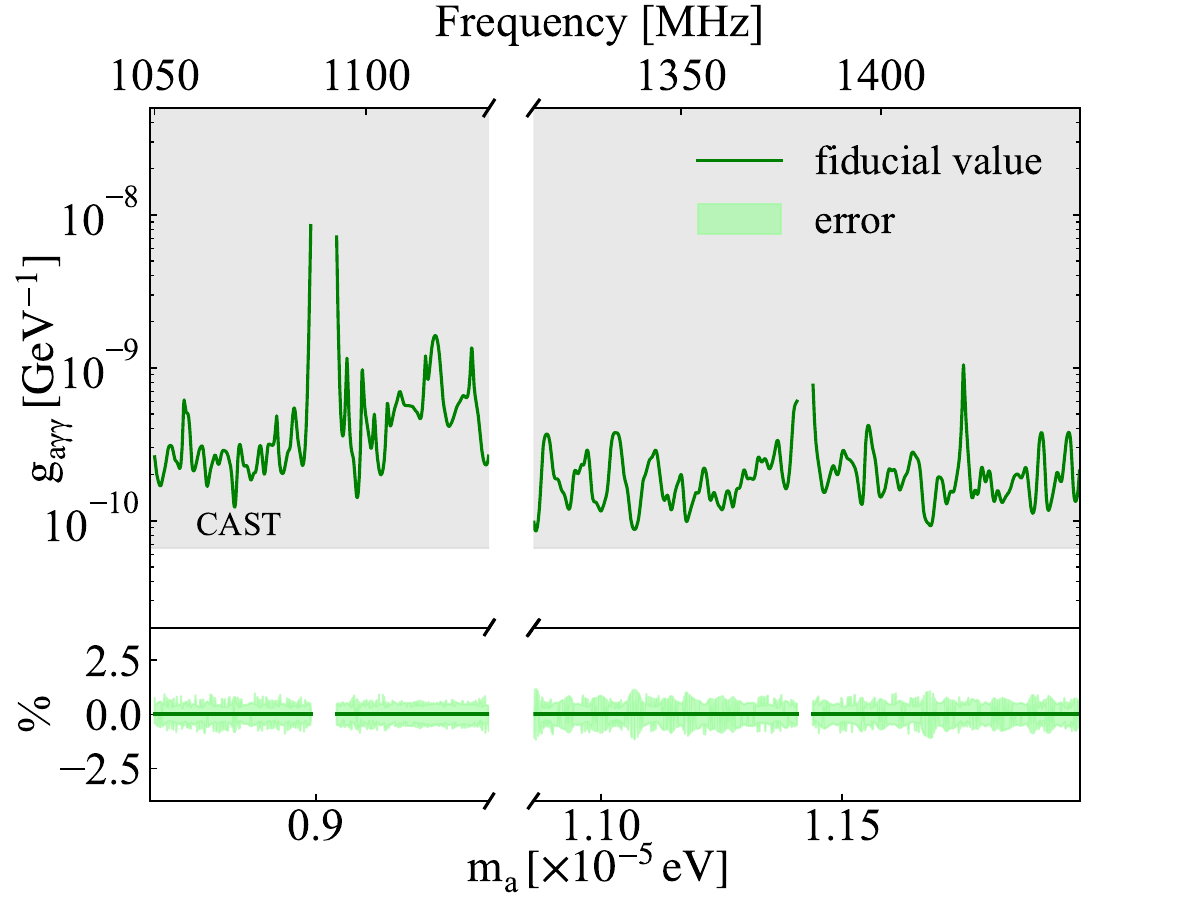}
    \caption{The error bar due to uncertainty in calibration process. The upper panel show absolute error while the lower panel shows the relative error. The solid lines represent our fiducial results and the colorful shadows mark the region of errors. The grey shadow shows result from CAST ($g_{a\gamma\gamma} \leq 6.6 \times 10^{-11} \mathrm{GeV}^{-1}$) \citep{2017NatPh..13..584A} for comparison. 
    }
    \label{fig:calibration_err}
\end{figure}

\subsection{Residual components} \label{subsec:discuss_residual}

\begin{figure*}
    \centering
    \includegraphics[width=0.9\textwidth]{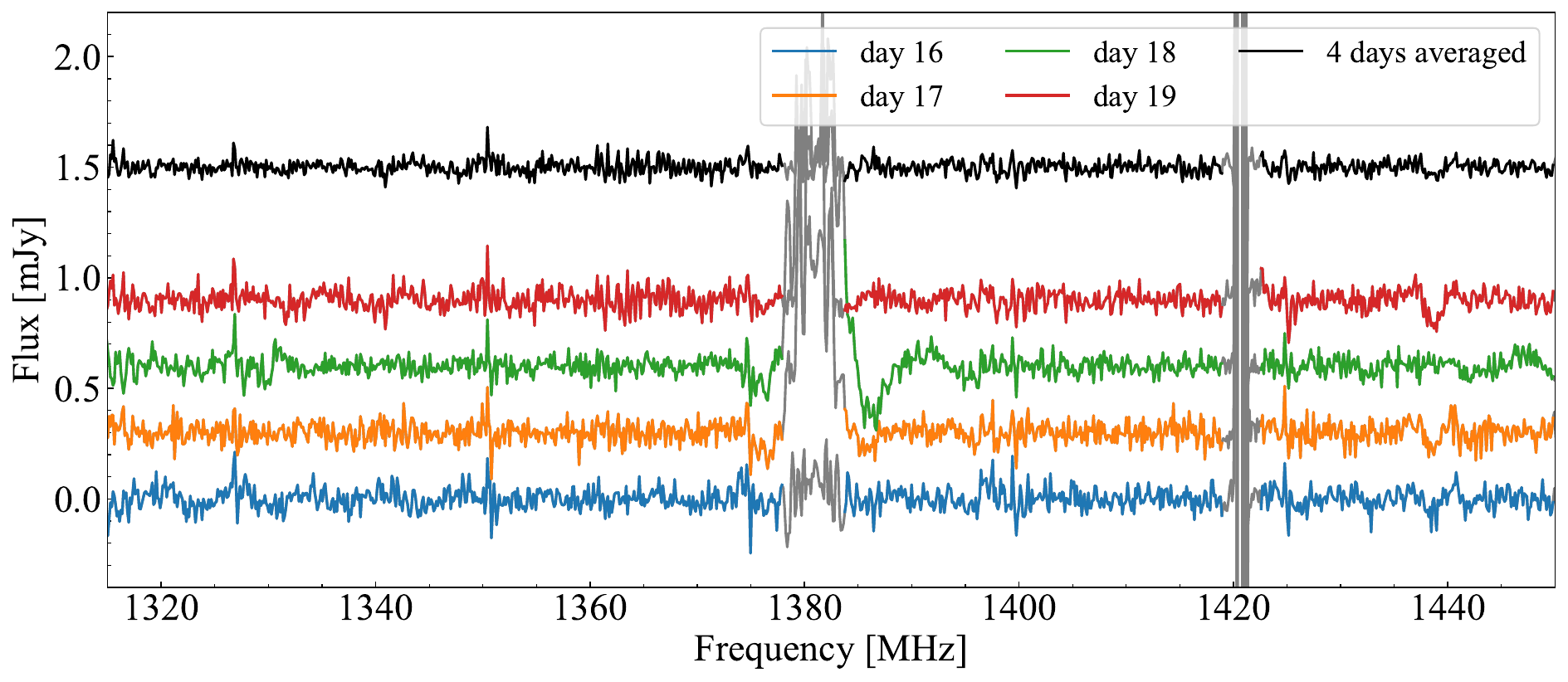}
    \caption{Spectra from 4 days for $\rm ON-OFF_3$ observation. The 4 colorful lines represent each day's spectrum respectively and the black line shows the spectrum for 4 days averaged data. Note that we move some spectra up to make all lines clear to see. All of them are actually centred at zero.} 
    \label{fig:spec_OFF3}
\end{figure*}

\begin{table} 
    \centering
    \caption{The rms of each spectrum in \reffg{fig:spec_OFF3}. }
     \label{tb:rms_OFF3}
    \begin{tabular}{cc}
        \\
        \hline
         \quad observation day \quad & \quad rms [$\mu\Jy$] \quad \\ 
         \hline
        \quad 16 \quad & \quad 46.3 \quad \\ 
        \quad 17 \quad & \quad 47.7 \quad \\ 
        \quad 18 \quad & \quad 50.3 \quad \\ 
        \quad 19 \quad & \quad 45.2 \quad \\ 
        \quad all 4 days \quad & \quad 30.0 \quad \\ 
         \hline
    \end{tabular}
\end{table}

\begin{figure}
    \centering
    \includegraphics[width=0.49\textwidth]{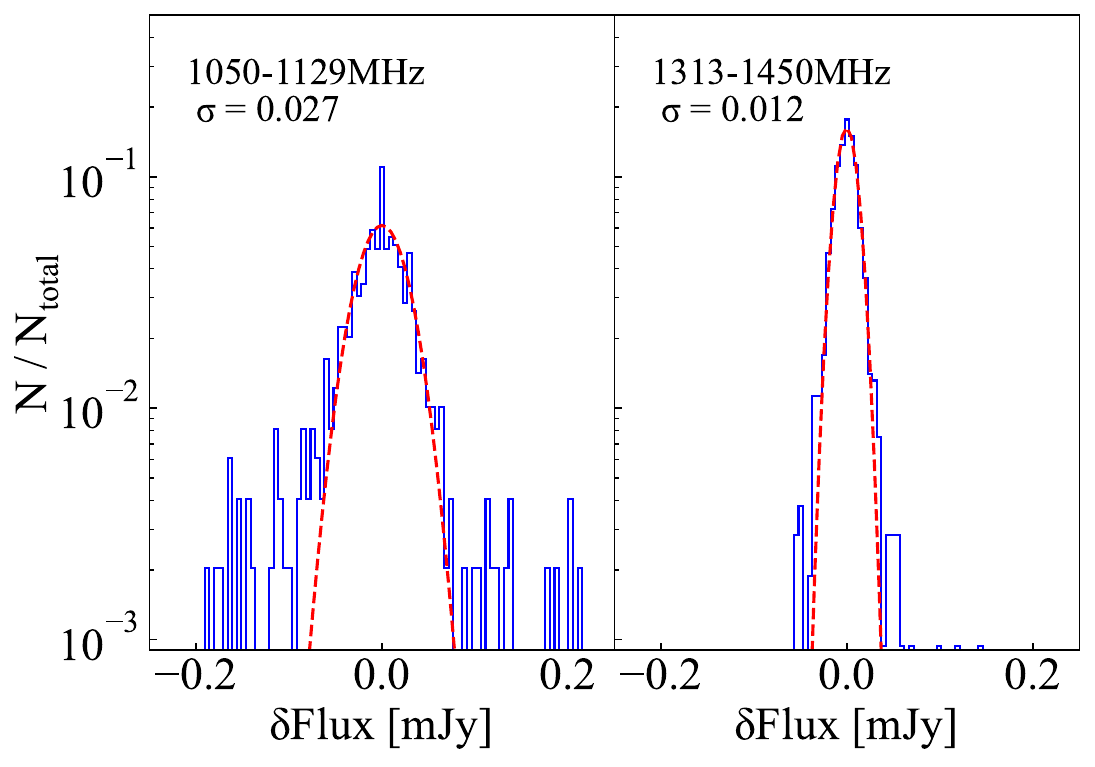}
    \caption{Histogram (blue lines) of noise in the final spectrum at frequency band 1050-1135MHz (left panel) and 1313-1450MHz (right panel). The red dashed lines show Gaussian fitted curve of the histogram. 
    }
    \label{fig:noise_gauss}
\end{figure}

\begin{figure}
    \centering
    \includegraphics[width=0.49\textwidth]{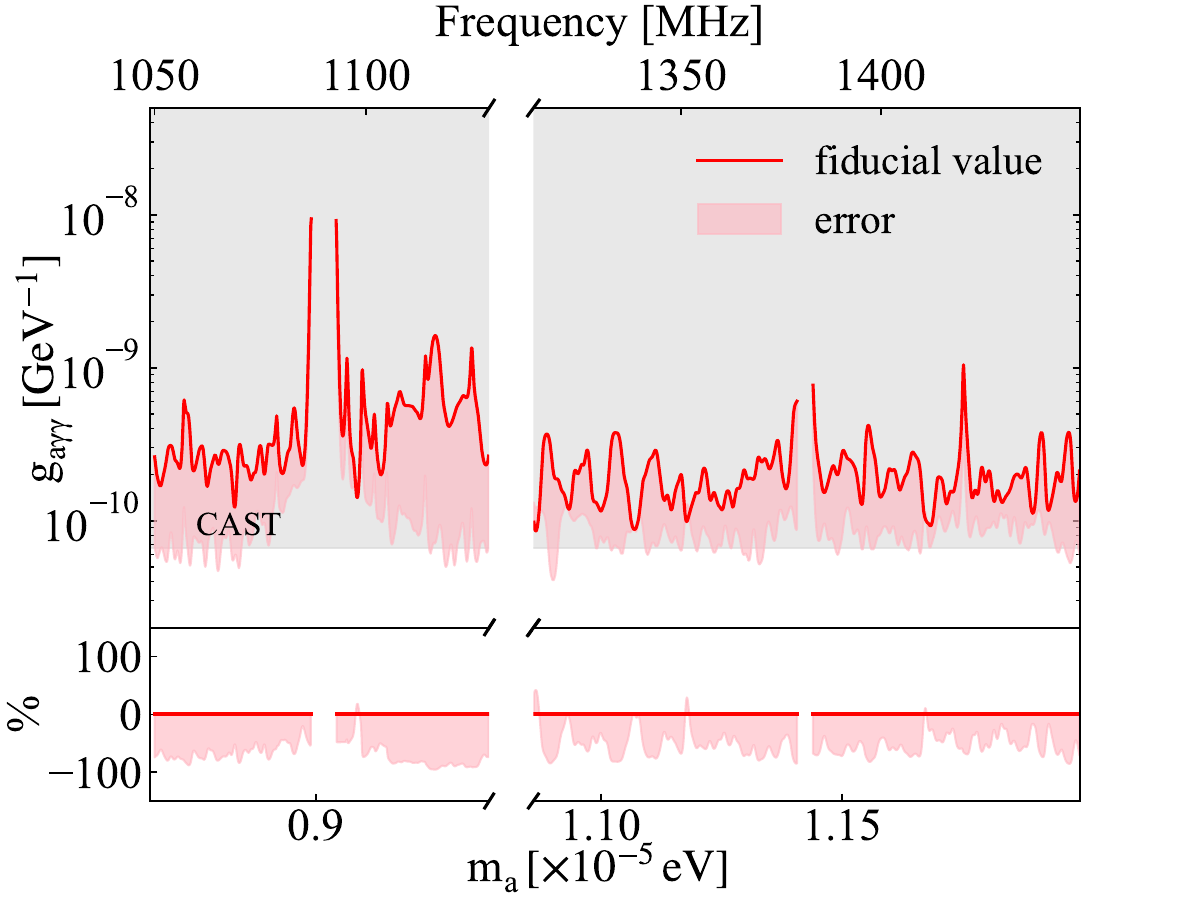}
    \caption{Similar to Fig.~\ref{fig:calibration_err}, but depicting the error  due to the residual RFI. }
    \label{fig:resi_err}
\end{figure}

Although strong temporal interference is removed through RFI flagging, there are inevitably residual RFIs. As our analysis is approaching sensitivity at the $\mu\Jy$ level, this residual RFI cannot be neglected.

We display four spectra obtained from 1 day's observation respectively and the averaged spectrum from these 4 days' observation in \reffg{fig:spec_OFF3} after baseline and standing waves removal. The rms values of each spectrum are listed in \reftb{tb:rms_OFF3}. 
The noise level does not decrease as much as we would expect according to \refeq{eq:sensitivity}, which confirms that the spectra contain more components beyond thermal noise. Possible origins of these extra components are RFIs, baseline residuals, standing wave residuals, $1/f$ noise \citealt{2021MNRAS.508.2897H, 2021MNRAS.501.4344L}, etc. Considering the risk of over-subtracting a real signal, it is necessary to be conservative 
while trying to avoid these extra components (e.g. not using 
too narrow a window for baseline subtraction to better fit the fluctuations). Therefore, similar to emission/absorption line signal detection, some better approaches for baseline/standing wave subtraction and RFI-contaminated data recovery may improve the results. 

To quantify the residual RFI and other possible sources of non-thermal noise, we make a histogram of the spectral noise in \reffg{fig:noise_gauss}. The noise should follow a Gaussian distribution if the RFI is successfully removed. We fit the histogram with a Gaussian function, and look for the residuals caused by the imperfect RFI removal. The non-Gaussian excess noise increases the rms by $\sim 35\%$ and $\sim8\%$ for the two frequency bands respectively. 

We also examine the effect of a stricter RFI-removing procedure. For each time session, we smooth the time-ordered calibrated data at each frequency point $d_{\nu_i}(t)$ over time by a median filter, and subtract the smoothed temporal baselines $b_{\nu_i}(t)$. The resulting difference, $n_{\nu_i}(t)= d_{\nu_i}(t) - b_{\nu_i}(t)$, is closer to a Gaussian distribution. 
We calculate its RMS and record $\mathrm{RMS}_{\nu_i}/\sqrt{t_{\nu_i, {\rm obs}}-t_{\nu_i,{\rm masked}}}$ as the projected error at $\nu_i$. However, the method to remove RFIs here is not suitable to be applied in our data processing part because a real signal will also be subtracted after this treatment, so we just use it estimate the ideal noise level if all RFIs are well removed.

Based on the noise spectrum derived above, we find that the effect of residual RFI contributes to an error depicted as a red shadow in \reffg{fig:resi_err}.
%}
The result is improved, reaching the result from CAST ($\sim 6.6 \times 10^{-11} \mathrm{GeV^{-1}}$) at some mass range, which is about 2 times better than the constraint shown in Fig.~\ref{fig:calibration_err}. Based on the size of this error, a reduction of residual RFI in the future could lead to a constraint on $g_{a\gamma\gamma}$ that is about twice as strong as the constraint shown in \reffg{fig:detect_limit}, which would be stronger than the CAST limit for some values of the axion mass.

\subsection{Pointing error} \label{subsec:discuss_pointing}

\begin{figure}
    \centering
    \includegraphics[width=0.45\textwidth]{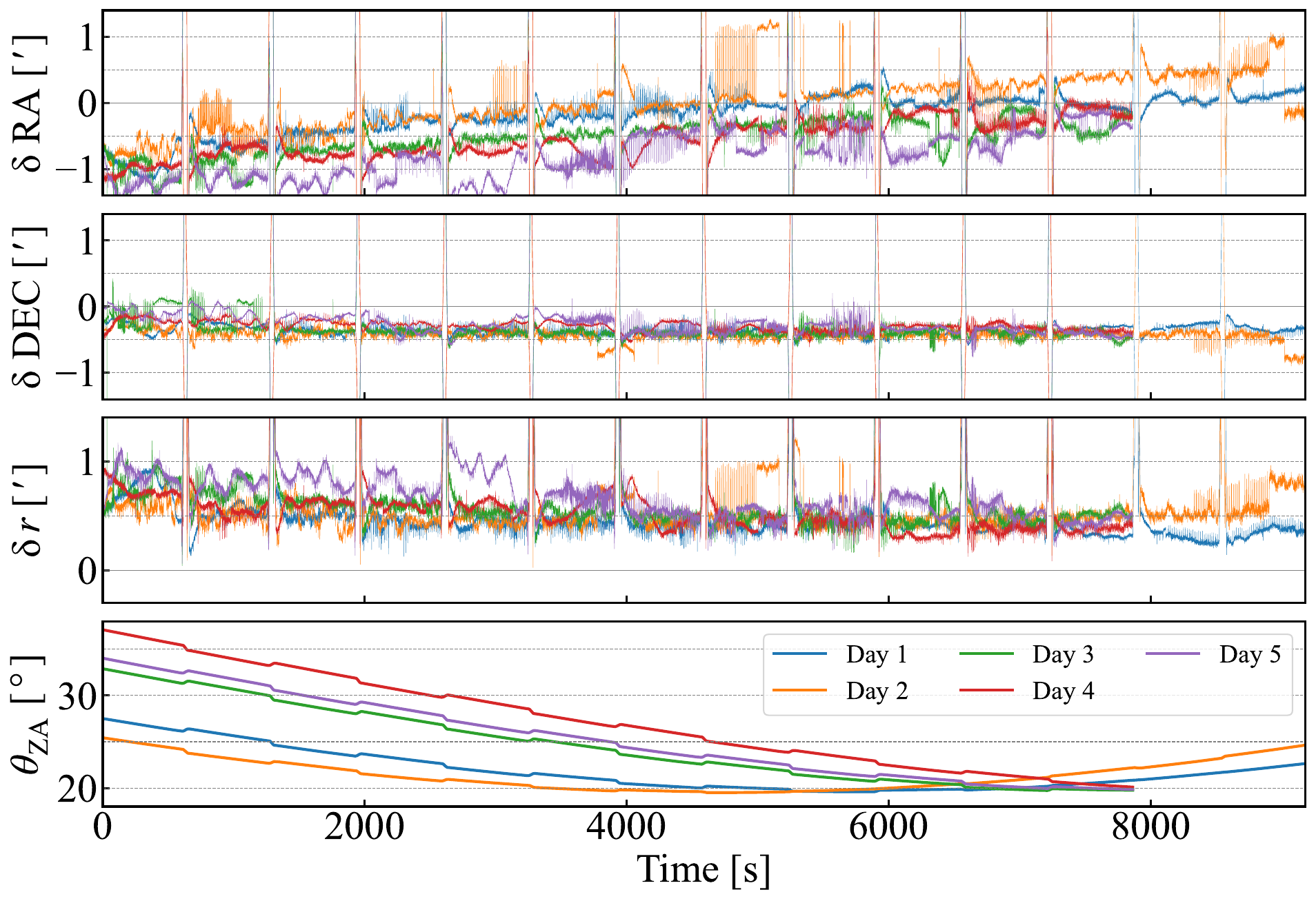}
    \caption{The pointing error in right ascension ($\delta$RA, top panel), and declination ($\delta$DEC, second panel), offset distance ($\delta$r, third panel) and zenith angle ($\theta_{ZA}$, bottom panel) during tracking observation in the 2022 observation season (B time). Different colors represent data from the 5 different days. }
    \label{fig:pointing}
\end{figure}

\begin{figure}
    \centering
    \includegraphics[width=0.47\textwidth]{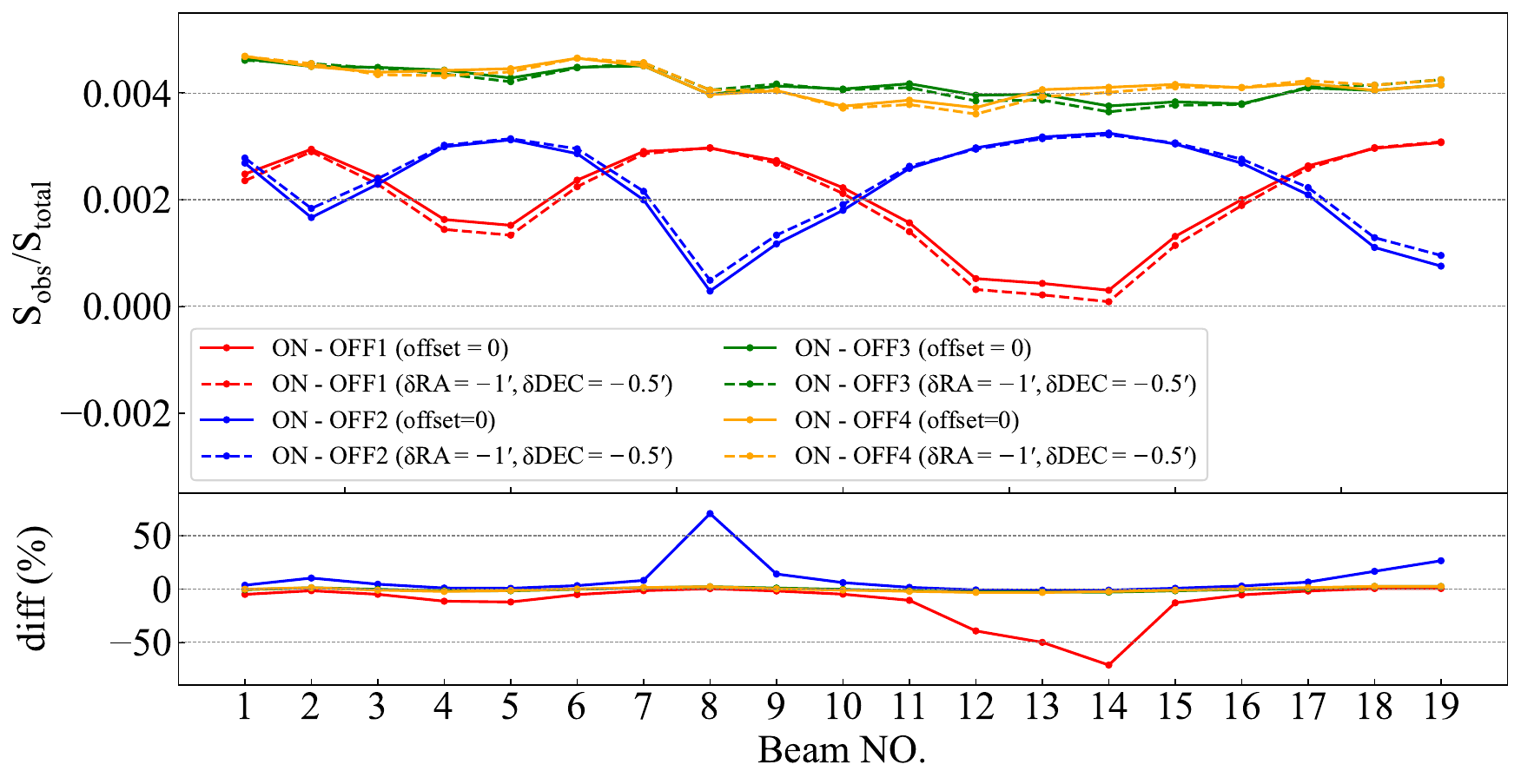}
    \caption{The simulated impact of pointing error ($\delta \mathrm{RA} = -1',\,\, \delta \mathrm{DEC} = -0.5'$) at 1250MHz. Upper panel: flux density of the 19 beams with (dashed lines) or without (solid lines) pointing deviation. Lower panel: relative variation in flux. Different colors represent the results for the 4 different OFF-source points. }
    \label{fig:dflux_offset}
\end{figure}

Although the telescope tracks the position of the gegenschein, its pointing is prone to some amount of error. As shown in \reffg{fig:pointing}, the deviation can reach up to $\sim$ 1 arcmin during the observation in one day. The deviation may also be correlated with the zenith angle, which is $20^{\circ} \sim 30^{\circ}$ in most days. We quantify the variation of the received signal induced by this effect through simulation. The gegenschein intensity is given \refeq{eq:Sg_vela}, then the observed flux with or without pointing error is calculated using \refeq{eq:S_obs} with the Gaussian beam model given in \citet{2020RAA....20...64J}. 
The variations of the target source flux (i.e. ON - OFF) at 1250~MHz observed by the 19 beams for an offset of $\delta \rm{RA} = -1~ \rm{arcmin}$ and $\delta \rm{DEC} = -0.5 ~\rm{arcmin}$ (which is close to the real offset) are shown in \reffg{fig:dflux_offset}. While using OFF3 or OFF4 for background subtraction, the influence of pointing error is negligible ($< 0.5\%$). However, the condition is more serious for OFF1 and OFF2. For a single beam, especially those at the outer ring, e.g. M14 in ON-OFF1, the flux variation could reach $\sim 70\%$. Nevertheless, the weighted averaging of the 19 beams data reduces the deviation to $\sim 3\%$ ($-3.27\%$ for OFF1 and $2.44 \%$ for OFF2 respectively), thanks to their symmetric distribution and the lower weights of beams at the outer ring. This variation is insensitive to frequency. The 1 arcmin deviation adopted here is a conservative estimation. The pointing error in most time is smaller than this value, typically $\sim 0.3-0.8$ arcmin. Here we assumed that the pointing measurements are precise enough that its error can be neglected. The estimated influence of pointing error to the constraint is therefore only about $1\sim2\%$.

\subsection{OFF-source points selection} \label{subsec:OFF}

The two OFF-source points used in data set A and B are fairly close to the expected gegenschein center, so there could be some gegenschein signal even at the OFF-source points. The loss of signal for the weighted spectral average of the 19 beams is estimated to be about 43\% for OFF1 and 40\% for OFF2, compared to ideal background subtraction. For the two further OFF-source position OFF3 and OFF4 used in data set C, the signal loss is significantly improved to be only about 2\%. 
However, although taking a more distant OFF-source position could reduce any residual gegenschein signal, it is also less reliable for sky background subtraction. This is especially true in this region of the sky which is close to the galactic plane, where galactic radiation can vary significantly. Hence, there is a balance between background subtraction and signal loss.

The observed spectra of the central beam (M01) at the four OFF-source positions are shown in \reffg{fig:spec_4OFF}, from which we can see a similarly undulating spectral shape and a disparity of 0.01~Jy$-$0.4~Jy between the curves. This is consistent across the data from all 19 beams. The difference between  observations pointing at the same position (the blue and orange curves in \reffg{fig:spec_4OFF}) is likely caused by system variations, such as the change of system temperature with zenith angle, which would be a few tenths of one Jy, as inferred from the measurement of 
\citet{2020RAA....20...64J}. 
The similar shape of the spectra at the different OFF-source positions (the orange and green curve in \reffg{fig:spec_4OFF}) indicates the similarity of the sky background. 

\begin{figure}
    \centering
    \includegraphics[width=0.45\textwidth]{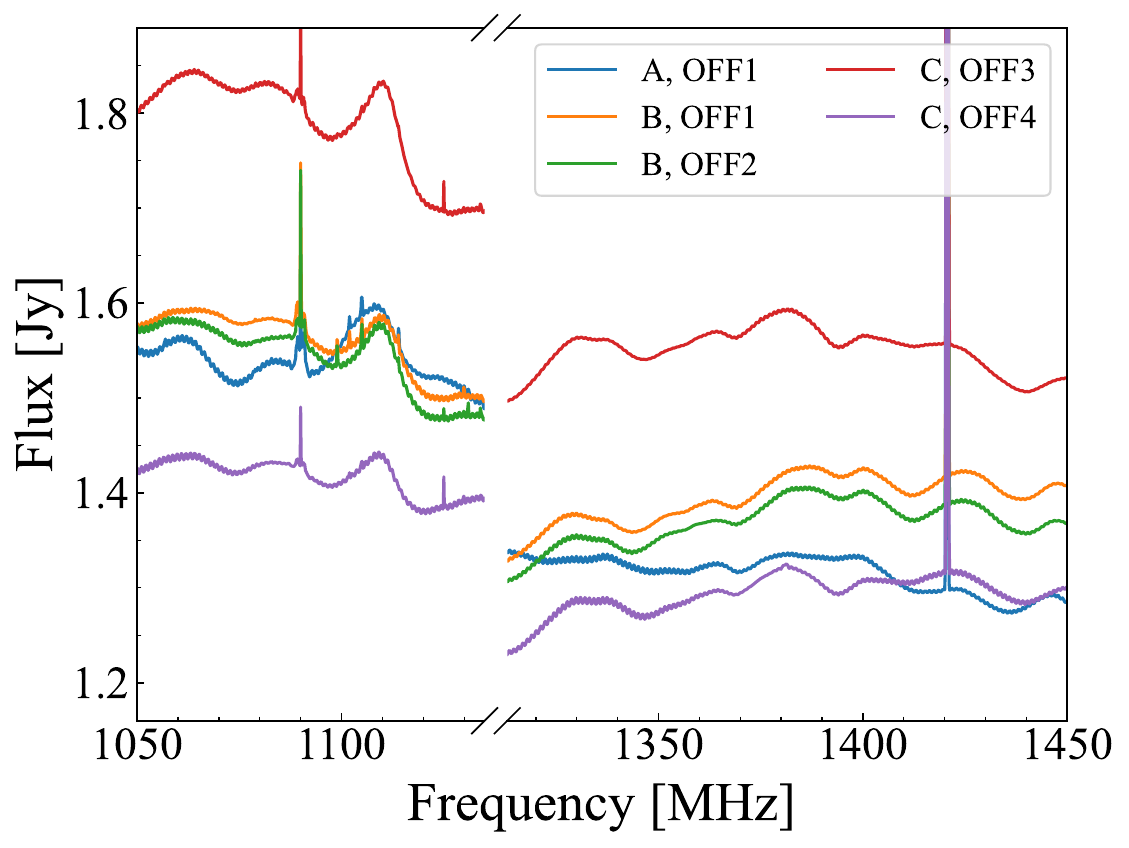}
    \caption{The flux spectra of different observations. The blue curve shows the spectrum at OFF-source position 1 in dataset A. The orange and green curve represent spectra at OFF-source positions 1 and 2 in dataset B. The red and purple lines are OFF-source position spectra in dataset C respectively. }
    \label{fig:spec_4OFF}
\end{figure}

\section{Summary} \label{sec:summary}

In this paper, we present the results of a pilot search for 
the axion gegenschein signal that would be induced by Vela SNR using FAST in the mass ranges $ 8.70\mu\mathrm{eV} \leq m_a \leq 9.44\mu\mathrm{eV}$ and $ 10.85\mu\mathrm{eV} \leq m_a \leq 12.01\mu\mathrm{eV}$. We search for the expected spectral line emission signal but observe no convincing candidate signal after performing statistical cross checks. 
The non-detection of an axion signal gives a constraint on the axion-photon coupling strength $g_{a\gamma\gamma}$, which we constrain to be smaller than $\sim 2\times 10^{-10} \mathrm{GeV}^{-1}$ in the two mass ranges we consider. Our analysis shows that non-negligible errors present in the data are mainly caused by residual components such as RFIs, baseline, and standing waves, while the influence of pointing error and calibration processes can be safely ignored. %through our detailed analysis. 
Our work has verified the feasibility of using the axion gegenschein technique, and the forecast result shows that the constraint is likely to be made tighter than CAST with $\sim$ 2000 hours ON-OFF mode observation in the future.

Axion gegenschein is a promising method for axion indirect detection at radio frequencies. % with many advantages. 
Compared with terrestrial experiments, these kinds of astrophysical approaches are more robust to any local non-uniformity of dark matter. The SNR gegenschein also gives a tighter constraint than other observable targets 
like dwarf galaxies or the galactic center because of the combined effects of the angular extent of the source 
and the dark matter density \citep{2020arXiv200802729G}. 

Future axion gegenschein searches can be extended and improved by observing additional sources (e.g. \citealt{2023arXiv231003788S}), more effective data processing methods for weak signal detection, and more integration time.

\section*{Acknowledgments}
This work made use of the data from FAST (Five-hundred-meter Aperture Spherical radio Telescope, \url{https://cstr.cn/31116.02.FAST}). FAST is a Chinese national mega-science facility, operated by National Astronomical Observatories, Chinese Academy of Sciences.
We acknowledge support from the  National SKA Program of China (Nos.2022SKA0110100 and 2022SKA0110101), the NSFC International (Regional) Cooperation and Exchange Project (No. 12361141814),
the CAS Interdisciplinary Innovation Team 
(JCTD-2019-05), and the science research grants from the China Manned Space Project with NO.CMS-CSST-2021-B01.
YL acknowledges the support of the National Natural Science Foundation of China (No. 1247309).

The authors thank Zheng Zheng, Jixia Li for helpful discussion.

\section*{Data Availability}

The data underlying this article will be shared on reasonable request to the corresponding authors.

% \appendix

% \setcounter{figure}{0}
% \setcounter{table}{0}
% \renewcommand{\thefigure}{A\arabic{figure}}
% \renewcommand{\thetable}{A\arabic{table}}

\bibliography{main}{}
\bibliographystyle{aasjournal}

\end{document}